\documentclass[fleqn,usenatbib]{mnras}
\usepackage[T1]{fontenc}

\DeclareRobustCommand{\VAN}[3]{#2}
\let\VANthebibliography\thebibliography
\def\thebibliography{\DeclareRobustCommand{\VAN}[3]{##3}\VANthebibliography}

\usepackage{graphicx}   
\usepackage{subcaption}
\usepackage{amsmath}    
\usepackage{amssymb}    
\usepackage{multicol}        
\usepackage{bm}     
\usepackage{pdflscape}  
\usepackage{multirow}
\usepackage{booktabs}
\usepackage{array}
\usepackage{float}
\usepackage{xspace} 
\usepackage{makecell}

\defcitealias{dMdB:2020}{dMdB20}

\newcommand{\Lya}{Ly-$\alpha$\xspace}

\newcommand{\beq}{\begin{equation}}
\newcommand{\eeq}{\end{equation}}
\newcommand{\beqa}{\begin{equation}\begin{aligned}}
\newcommand{\eeqa}{\end{aligned}\end{equation}}

\newcommand{\vk}{\mathbf{k}}

\newcommand{\vr}{\mathbf{r}}

\newcommand{\td}{\delta}
\newcommand{\dd}{\partial}
\newcommand{\kvec}{\mathbf{k}}

\newcommand{\ap}{\alpha_{\parallel}}
\newcommand{\at}{\alpha_{\perp}}

\newcommand{\kpar}{k_{\parallel}}
\newcommand{\kperp}{k_{\perp}}

\newcommand{\kvperp}{\mathbf{k}_{\perp}}
\newcommand{\rpar}{r_{\parallel}}
\newcommand{\rperp}{r_{\perp}}

\newcommand{\rvperp}{\mathbf{r}_{\perp}}

\newcommand{\mat}[1]{\mathbf{#1}}
\newcommand*\diff{\mathop{}\!\mathrm{d}}

\newcommand*{\eg}{e.g.\xspace}
\newcommand*{\ie}{\emph{i.e.}\xspace}
\newcommand{\lamobs}{\lambda_{\mathrm{obs}}}
\newcommand{\lamrf}{\lambda_{\mathrm{rf}}}

\newcommand{\hinvMpc}{\,h^{-1}\, {\rm Mpc}\,}
\newcommand{\Hunit}{\, {\rm km}\,s^{-1}\, {\rm Mpc}^{-1}\,}
\newcommand{\hMpcinv}{\,h\, {\rm Mpc}^{-1}\,}
\newcommand{\hinvGpc}{\,h^{-1}\, {\rm Gpc}\,}

\newcommand{\phiinv}{\Phi^{-1}}

\newcommand{\hi}{H\,{\sc i}~}

\newcommand{\siii}{Si\,{\sc ii}~}
\newcommand{\siiii}{Si\,{\sc iii}~}

\def\ltsima{$\; \buildrel < \over \sim \;$}
\def\gtsima{$\; \buildrel > \over \sim \;$}
\def\simlt{\lower.5ex\hbox{\ltsima}}
\def\simgt{\lower.5ex\hbox{\gtsima}}

\newcommand{\Lyacolore}{\texttt{LyaCoLoRe}\xspace}
\newcommand{\Planck}{\emph{Planck}\xspace}

\providecommand{\sorthelp}[1]{}

\newcolumntype{N}{>{\centering\arraybackslash}m{.5in}}
\newcolumntype{G}{>{\centering\arraybackslash}m{2in}}

\usepackage{newtxtext,newtxmath}

\title[3D \Lya Forest Power Spectrum]{The 3D Lyman-$\alpha$ Forest Power Spectrum from eBOSS DR16}

\author[de Belsunce et al.]{Roger de Belsunce,$^{1,2}$\thanks{\href{mailto:rbelsunce@berkeley.edu}{rbelsunce@berkeley.edu}} 
Oliver H.~E.~Philcox,$^{3,4}$\thanks{\href{mailto:ohep2@cantab.ac.uk}{ohep2@cantab.ac.uk}} 
Vid Ir\v{s}i\v{c},$^{5,6}$ 
Patrick McDonald,$^{1}$%
\newauthor
Julien Guy,$^{1}$
and Nathalie Palanque-Delabrouille$^{1}$
\\
$^{1}$Lawrence Berkeley National Laboratory, One Cyclotron Road, Berkeley CA 94720, USA\\
$^{2}$Berkeley Center for Cosmological Physics, Department of Physics, University of California, Berkeley, CA 94720, USA\\
$^{3}$Department of Physics, Columbia University, New York, NY 10027, USA\\
$^{4}$Simons Society of Fellows, Simons Foundation, New York, NY 10010, USA\\
$^{5}$Kavli Institute for Cosmology, University of Cambridge, Madingley Road, Cambridge CB3 OHA, UK\\
$^{6}$Cavendish Laboratory, University of Cambridge, 19 J. J. Thomson Ave., Cambridge CB3 0HE, UK
}

\date{Accepted XXX. Received YYY; in original form ZZZ}

\pubyear{2024}

\begin{document}
\label{firstpage}
\pagerange{\pageref{firstpage}--\pageref{lastpage}}
\maketitle

\begin{abstract}
We measure the three-dimensional power spectrum (P3D) of the transmitted flux in the Lyman-$\alpha$ (\Lya) forest using the complete extended Baryon Oscillation Spectroscopic Survey data release 16 (eBOSS DR16). This sample consists of $\sim$\,$205,000$ quasar spectra in the redshift range $2\leq z \leq 4$ at an effective redshift $z=2.334$. We propose a pair-count spectral estimator in configuration space, weighting each pair by $\exp(i\mat{k}\cdot \mat{r})$, for wave vector  $\mat{k}$ and pixel pair separation $\mat r$, effectively measuring the anisotropic power spectrum without the need for fast Fourier transforms. This accounts for the window matrix in a tractable way, avoiding artifacts found in Fourier-transform based power spectrum estimators due to the sparse sampling transverse to the line-of-sight of \Lya skewers. We extensively test our pipeline on two sets of mocks: (i) idealized Gaussian random fields with a sparse sampling of \Lya skewers, and (ii) log-normal \Lyacolore mocks including realistic noise levels, the eBOSS survey geometry and contaminants. On eBOSS DR16 data, the Kaiser formula with a non-linear correction term obtained from hydrodynamic simulations yields a good fit to the power spectrum data in the range $(0.02 \leq k \leq 0.35)\hMpcinv$ at the $1-2\sigma$ level with a covariance matrix derived from \Lyacolore mocks. We demonstrate a promising new approach for full-shape cosmological analyses of \Lya forest data from cosmological surveys such as eBOSS, the currently observing Dark Energy Spectroscopic Instrument and future surveys such as the Prime Focus Spectrograph, WEAVE-QSO and 4MOST. 
\end{abstract}


\section{Introduction} \label{sec:intro}
Light is absorbed by neutral hydrogen in the low-density, highly ionized intergalactic medium (IGM), producing a series of characteristic absorption features in quasar spectra, called the Lyman-$\alpha$ (\Lya) forest. These absorption patterns are a measure of the distribution of neutral hydrogen along the line-of-sight. They represent a fundamental probe of large-scale structure at Mpc scales and below at high redshifts ($2 \simlt z \simlt 5$) using ground-based observations.  Given the sparseness of the \Lya forest\footnote{The \Lya forest is sparsely sampled transverse to the line-of-sight but densely sampled along the line-of-sight.}, the two-point correlation function \citep{Slosar2011} and the one-dimensional power spectrum (P1D) along the line-of-sight  \citep{McDonald06} have been widely adopted in cosmological analyses of \Lya data \citep[e.g.,][]{Slosar2013, PYB13, Busca:2013, dMdB:2020}. 

The two-point correlation function (2PCF) can efficiently isolate effects at specific separations, \eg, the baryonic acoustic oscillation (BAO) feature, whilst the power spectrum is a more convenient statistic for measuring slowly varying effects down to small wavenumbers and provides less correlated errors \citep{Font-Ribera:2018}. In this work, we propose a weighted pair-count estimator, developed for the analysis of small-scale galaxy clustering \citep{Philcox2020, Philcox:2021}, to measure the 3D \Lya forest power spectrum (P3D). By weighting each pair of \Lya pixels by $\exp{(i\mat{k}\cdot \mat{r}_{ij})}$ (for wave vector $\mat{k}$ and pair separation $\mat{r}_{ij}\equiv\mat{r}_{i}-\mat{r}_{j}$), we directly measure the power spectrum without the need for grid-based fast Fourier-transforms (FFT) which are strongly affected by the highly non-trivial window function specific to \Lya surveys. This method also allows for almost exact computation of the window function using the same pixel pairs, albeit at a significant computational cost of $\mathcal{O}(N_{\rm pix}^2R_0^3)$ (for number of spectral pixels $N_{\rm pix}$ and maximum pair separation $R_0$).

Over the past decades, \Lya surveys have improved in accuracy, size, and depth, resulting in large samples of medium-resolution spectra from the extended Baryon Oscillation Spectroscopic Survey \citep[eBOSS;][]{Dawson:2016} and the currently observing Dark Energy Spectroscopic Instrument \citep[DESI;][]{DESI:2016}. Whilst the eBOSS DR16 sample consists of $205,012$ quasar spectra with absorber redshifts in the range $1.96 \leq z \leq 3.93$, DESI observes the sky with increased spectral resolution of $2000 \simlt \mathrm{R}\simlt 5500$ \citep{DESI:2022} and covers a much larger fraction of the sky (14,000 deg$^2$). Over its survey duration it will provide $\sim$\,$840,000$ quasar spectra at $z>2.1$, a more than fourfold increase compared to BOSS/eBOSS \citep{2024AJ....167...62A}. To complement the picture, smaller samples of high-resolution measurements from, e.g., the High Resolution Echelle Spectrometer \citep[HIRES;][]{Vogt:1994,OMeara:2021} and the UV-Visual Echelle Spectrograph \citep[UVES;][]{Dekker:2000,Murphy:2019} allow for analyses deeper into the small-scale regime to almost $10x$ larger $k_{\rm max}$. A number of future surveys will also capture spectra in both high and medium resolution modes, such as the WEAVE-QSO survey \citep{2016sf2a.conf..259P}, the Prime Focus Spectrograph \citep[PFS;][]{2022PFSGE} and 4MOST \citep{2019Msngr.175....3D}. A fast and robust estimator for the analysis of \Lya forest at all scales is essential for accessing the full potential of the combination of these data sets. 

One of the key advantages of the \Lya forest is that, since the density fields are only mildly non-linear at the respective redshifts, a much wider range of scales can be used to robustly probe cosmology than with most galaxy surveys. This makes the \Lya fluctuations a particularly powerful probe of early-Universe physics when combined with tracers that are sensitive to very large scales, e.g,~cosmic microwave background anisotropies. The \Lya flux density contrast, $\delta_F$, is used to derive constraints for cosmology and astrophysics, defined through
\begin{equation} \label{eq:delta_F}
    \delta_F(\lambda) = \frac{f_q(\lambda)}{\overline{F}(z_{\text{\Lya}})q(\lambda)}-1\ ,
\end{equation} 
where $f_q(\lambda)$ is the observed transmitted flux as a function of $\tau$, the optical depth, $\overline{F}(z_{\text{\Lya}})$ is the mean transmitted flux at the HI absorber redshift and $q(\lambda)$ is the unabsorbed continuum of the background quasar. The three-dimensional power spectrum, $P_{\rm 3D}$, of the fluctuations in the \Lya forest, given in Eq.~\eqref{eq:delta_F}, is principally sensitive to the amplitude of dark matter clustering $\sigma_8$, the shape of the matter power spectrum $\Gamma=\Omega_{\rm m}h$, the spectral tilt $n_s$ and the sum of neutrino masses $\sum m_{\nu}$ \citep[see, e.g.,~][]{2022PhRvD.105d3517P}.

In this work, we present a measurement of the anisotropic 3D \Lya forest power spectrum using eBOSS DR16 \Lya forest spectra. We compare our measurement to the best-fit results from the 2PCF analysis in the range $(0.019 \leq k \leq 0.35)\hMpcinv$ \citep[][hereafter \citetalias{dMdB:2020}]{dMdB:2020}. It includes a Kaiser formula \citep{Kaiser1987} with a non-linear correction term obtained from hydrodynamical simulations \citep{Arinyo:2015} and modeling of contaminants (e.g.,~metals, distortion from continuum fitting, damped \Lya absorbers) specific to \Lya forest analyses. Predictions for the \Lya forest clustering have greatly improved over the last decade, using either hydrodynamical simulations \citep[see, e.g.,][]{Arinyo:2015, Bolton:2017, 2023MNRAS.519.6162P, 2023arXiv231201480D} or perturbation theory\footnote{On analyses of the connection between the physics of the \Lya forest and perturbative approaches, \ie, response function approaches, see \cite{Seljak:2012, Cieplak:2016, Irsic:2018}.} \citep[see, e.g.,][]{Givans:2020, Garny:2021, Chen:2021, Givans:2022, Ivanov:2023}.

In the context of simulations, different P3D methods have been proposed and tested, see~\cite{McDonald:2001, Arinyo:2015, Font-Ribera:2018, Horowitz:2024}. The latter two methods provide close-to-optimal ways of constructing a covariance matrix. Whilst we measure the anisotropic clustering in multipole space and as a function of $k=\sqrt{\kpar^2+\kvperp^2}$, \citet{Font-Ribera:2018} measure the P3D in the (more natural basis) $\{\kpar, \kvperp\}$-basis (which is the Fourier space equivalent to the $\{\rpar, \rvperp\}$-basis of the configuration space analysis in, e.g.,~\citetalias{dMdB:2020}). Whilst the information content is (in principle) the same as in the present analysis, the $\{\kpar, \kvperp\}$-basis allows to isolate (and marginalize out) modes that are, e.g.,~ sensitive to distortions in the continuum fitting ($\kpar =0$) \citep[see][for a discussion]{Font-Ribera:2018}. In the present work we compute a simulation-based covariance matrix, whereas \cite{Font-Ribera:2018} compute an approximate version of the global covariance matrix within the optimal quadratic estimator framework. A recent proof-of-principle presented in \cite{karim2023measurement} presents a small-scale measurement on simulations and eBOSS DR16 data which measures the cross-spectrum $P_{\times}(z,\theta,\kpar)$, originally developed in \cite{Hui:1999, Font-Ribera:2018}.

The \Lya forest is a treasure trove of cosmological information, capable of measuring the expansion history of the Universe through baryonic acoustic oscillations \citep[BAO;][]{McDonald:2007,Slosar2013, Busca:2013, dMdB:2020}, the broadband shape of the (large-scale) 3D \Lya correlation function \citep{Slosar2013, Cuceu:2021, Cuceu:2023, Gordon:2023}, neutrino masses \citep{Seljak:2005, Viel:2010, PYB13, Palanque2020}, and primordial black holes \citep{Afshordi:2003, Murgia:2019}. The P3D is connected to the well-understood one-dimensional power spectrum (P1D) by performing an integration perpendicular to the line-of-sight, which is sensitive to small scales \citep[see, e.g.,][]{Seljak:2005, Viel:2005,McDonald06, PYB13, Chabanier:2019, Pedersen:2020,  2023MNRAS.526.5118R, 2024MNRAS.tmp..176K}. At the smallest of scales, the \Lya absorption features provide means to test dark matter models \citep{Viel:2013, Baur:2016, Irsic17, Kobayashi:2017, Armengaud:2017, Murgia:2018,Garzilli:2019, Irsic:2020, Rogers:2022, Villasenor:2023, Irsic:2023}, early dark energy models \citep{2023PhRvL.131t1001G}, and thermal properties of the ionized (cold) IGM \citep{Zaldarriaga:2002, Meiksin:2009,McQuinn:2016, Viel:2006, Walther:2019, Bolton:2008, Garzilli:2012, Gaikwad:2019, Boera:2019, Gaikwad:2021, Wilson:2022, Villasenor:2022}.

The remainder of this paper is organized as follows: We present the 3D power spectrum estimator in Sec.~\ref{sec:methods}, before discussing the theoretical modeling of the \Lya flux power spectrum in Sec.~\ref{sec:theory_pk} and the forward modeling of the window function in Sec.~\ref{Sec:window_func}. In Sec.~\ref{sec:data} we describe the observational eBOSS DR16 \Lya forest data set as well as the employed synthetic \Lya spectra used to model covariances. In Sec.~\ref{sec:simulations} we extensively test our pipeline on mocks before presenting the main result of our analysis in Sec.~\ref{sec:results}: the eBOSS DR16 three-dimensional power spectrum of the \Lya forest. Sec.~\ref{sec:conclusions} presents our conclusions. We make our \texttt{HIPSTER-lya} implementation publicly available.\footnote{\url{https://github.com/oliverphilcox/HIPSTER}}

\section{The 3D power spectrum estimator} \label{sec:methods}
In cosmological data analysis, one typically employs data compression, both to reduce the size of the data vector to a computationally manageable level and average over stochastic fluctuations. The particular compression statistic is specific to the problem at hand; for \Lya, one typically uses the 2PCF or the one-dimensional power spectrum along the line-of-sight \citep[e.g.,][]{Slosar2013, Busca:2013, PYB13, 2023MNRAS.526.5118R, 2024MNRAS.tmp..176K}. Here, we compress the large number of \Lya forest spectra instead into the 3D power spectrum,\footnote{Throughout the paper, we use the following notation: $\Hat{P}(k)$ are 3D power spectra measured on the data. Theory spectra are denoted by $P(k)$ which, when convolved with the window matrix, are labeled $\Tilde{P}(k)$. The multipoles of the power spectra are denoted by the subscript $\ell$, \ie, $P_{\ell}(k)$.} paving the way for future full-shape cosmological analyses in Fourier space of eBOSS data (and beyond).\footnote{For a compressed full-shape analysis in configuration space, see, e.g.,~\cite{Cuceu:2023}.}

Usually, \Lya summary statistics are computed using quadratic maximum likelihood \citep[e.g.,][]{McDonald06, NaimQMLE2020} or FFT-based algorithms \citep[e.g.,][]{PYB13, Chabanier:2019}; in this work, we instead make use of configuration space pair-count estimators, following the \texttt{HIPSTER} algorithm of \citep{Philcox2020, Philcox:2021}. These were originally introduced in the context of the small-scale galaxy power spectrum and bispectrum, and translate well to our problem, given the sparse (dense) sampling of pixels transverse to (along) the line-of-sight, which typically leads to aliasing for FFT-based methods. As we discuss below, \texttt{HIPSTER} estimates spectra by summing pairs of points in configuration space, weighted by $\exp{(i\mat{k}\cdot \mat{r}_{ij})}$, for wave vector $\mat{k}$ and pair separation by $\mat{r}_{ij}\equiv\mat{r}_{i}-\mat{r}_{j}$. 

Given some data field $D(\mat{x})$ (encoding \Lya fluctuations or galaxy overdensities, with some weighting), the (pseudo-)power spectrum is defined as a Fourier-transform of the two-point function $DD$:
\begin{equation} \label{eq:FFT_2PCF}
    P(\mat k) \equiv \mathcal{F}\left[ DD(\mat r) \right] = \int \mathrm{d}^3 \mat r\  DD(\mat r)e^{i\mat{k}\cdot \mat{r}} \ ,
\end{equation}
where $\mathcal{F}$ is the Fourier transform operator, and we have assumed $DD$ to be suitably normalized. Usually, one considers angularly-averaged power spectra, defined by\footnote{We use the shorthand notation: $\int_{\mat x}$ for $\int \mathrm{d}^3\mat{x}$ and $\int\displaylimits_{\mat{k}}$ for $\int \mathrm{d}^3\vk$.}
\begin{equation}\label{eq: k-binning}
P^{a} = \frac{1}{V_a} \int\displaylimits_{\mat{k}}\Theta^{a}(|\mat{k}|)P(\mat k) \ , 
\end{equation}
where we have restricted to a $k$-bin $a$, with volume $V_a=\int\displaylimits_{\mat{k}}\Theta^{a}(|\mat{k}|)$, for top-hat function $\Theta^a(k)$. Comparing to Eq.~\eqref{eq:FFT_2PCF}, we see that the $\mat{k}$-dependence is captured through the kernel 
\begin{equation}
    A^{a}(\mat{r}_i-\mat{r}_j) \equiv \frac{1}{V_a}\int\displaylimits_{\mat{k}} \Theta^{a}(k) e^{i\mat{k}\cdot (\mat{r}_i-\mat{r}_j)} \approx j_0(k_a|\mat{r}_i-\mat{r}_j|) \ ,
\end{equation}
where $j_0$ is the 0-th order spherical Bessel function. In the last step, we assumed the thin-shell limit and aligned the $|\Hat{\mat{k}}|$-axis with $|\mat{r}_i-\mat{r}_j|$. Noting that $D(\mat x)=n(\mat x)w(\mat x)\delta(\mat x)$ for background density $n$, weights $w$, and overdensity $\delta$ (which is the quantity of interest), the power spectrum can be written explicitly as
\begin{align}\label{eq: P-iso-a}
    P^{a} & = \frac{1}{\overline{V(nw)^2}} \int\displaylimits_{\mat{r}_i}\int\displaylimits_{\mat{r}_j}n(\mat r_i)n(\mat r_j) w(\mat r_i)w(\mat r_j)\delta(\mat r_i)\delta(\mat r_j) \\
    & \times A^{a}(\mat{r}_i-\mat{r}_j) \ , \nonumber
\end{align}
adding a normalization term appropriate to an ideal uniform survey of volume $V$. In this approach, the power spectrum can be computed by explicitly evaluating Eq.~\eqref{eq: P-iso-a}, \textit{i.e.}\ counting each pair of \Lya pixels, weighted by above kernel. 

In practice, we also wish to estimate the anisotropy of the \Lya power spectrum. This is possible by a simple generalization for multipole $\ell$:
\begin{align}\label{eq:P-aniso-a}
    P^{a}_\ell & = \frac{1}{\overline{V(nw)^2}} \int\displaylimits_{\mat{r}_i}\int\displaylimits_{\mat{r}_j}n(\mat r_i)n(\mat r_j) w(\mat r_i)w(\mat r_j)\delta(\mat r_i)\delta(\mat r_j) \\
    & \times A^{a}_{\ell}(\mat{r}_i,\mat{r}_j)\nonumber\\
    A_{\ell}^{a}(\mat r_i, \mat r_j) &\equiv \frac{2\ell +1}{V_a}\int\displaylimits_{\mat{k}} \Theta^{a}(|\mat k|) e^{i\mat{k}\cdot(\mat{r}_i-\mat{r}_j)} \mathcal{L}_{\ell}(\measuredangle[\mat k, \frac{1}{2}(\mat r_i+\mat r_j)]) \nonumber \ ,
\end{align}
where $\measuredangle[\mat k, \frac{1}{2}(\mat r_i+\mat r_j)]$ denotes the cosine of the angle between $\vk$ and the local line-of-sight $(\vr_i+\vr_j)/2$. Eq.~\eqref{eq:P-aniso-a} is the basis of the three dimensional power spectrum estimator used herein\footnote{In \citet{Philcox2020}, an additional $\phiinv$ term was included to approximately deconvolve the effects of the survey geometry. In this work, we instead forward model the effects of the window (to avoid instabilities), as discussed in Sec.~\ref{Sec:window_func}.}.

In practice, all pairs of \Lya pixels contribute to the measured power spectrum, which results in the na\"ive estimator of Eq.~\eqref{eq:P-aniso-a} being extremely expensive to compute. In practice however, well-separated pairs have negligible contributions to the power spectrum, thus we can truncate the pair counts at separations greater than $R_0 \geq 200 \hinvMpc$, with minimal loss of information \citep{Philcox2020}. This is done via a smooth polynomial window function 
\begin{equation} \label{eq:wofr}
    W\left( \frac{\mat{|r|}}{R_0}\right)  \equiv
\begin{cases}
    1,&   0\leq \frac{r}{R_0} < \frac{1}{2} \\
    1-8(2\frac{r}{R_0}-1)^3 + 8(2\frac{r}{R_0}-1)^4,&  \frac{1}{2}\leq \frac{r}{R_0} < \frac{3}{4} \\
    -64(\frac{r}{R_0}-1)^3 - 128(\frac{r}{R_0}-1)^4,&  \frac{3}{4} \leq \frac{r}{R_0} < 1 \\
    0,&  \text{else},
\end{cases}
\end{equation}
which modifies the anisotropic power spectrum estimator in Eq.~\eqref{eq:P-aniso-a} by a factor $W((\mat{r}_i-\mat{r}_j)/R_0)$. Though this necessarily induces slight distortions to the measured spectra, its effect can be straightforwardly included in the theoretical model for $P_\ell^a$, thus avoiding any potential bias.

Finally, we must consider the window function of the data, denoted by $\Phi(\mat r) \equiv DD(\mat r)/\xi(\mat r)$, where $\xi(\mat r)=\langle{\delta(\mat x)\delta(\mat x+\mat r)\rangle}$, is the true 2PCF of the data. Due to the sampling of the \Lya skewers, this has a non-trivial form, and leads to our estimators returning a window-convolved power spectrum measurement. Explicitly, this effect can be included by forward modeling the window function in the theoretical model, using
\begin{align}\label{eq: pk-conv}
\Tilde{P}(\mat k) &= \mathcal{F}\left[\xi(\mat r) \Phi(\mat r)W(r;R_0)\right](\mat k)\ ,
\end{align}
where $\xi(\mat r)$ is the correlation function model, and $\Tilde{P}(\mat k)$ is the output window- (and pair-separation-)convolved power. This procedure will be discussed in more detail in Sec.~\ref{Sec:window_func}.

In discrete form, the full estimator for the \Lya power spectrum used in this work is given by a sum over each pair of pixels, $i,j$:
\begin{equation} \label{eq:DD_pk}
    \hat{P}^a_\ell =\frac{1}{\overline{V(nw)^2}}\sum_{i}\sum_{j,\, i \neq j} w_iw_j \td_{F,i}\td_{F,j} A^{a}_\ell(\mat{r}_i,\mat{r}_j) W(|\mat{r}_i-\mat{r}_j|; R_0) \ ,
\end{equation}
where we drop the (arbitrary) normalization factor and additionally remove the `self-skewer' counts; an advantage of doing the measurement in configuration space to avoid (correlated) uncertainties stemming from continuum estimation. The weights $w$ are obtained from the continuum fitting \citepalias{dMdB:2020}, and $\delta_{F,i}$ are the estimated \Lya fluctuations. We use the same set of pixels to compute the window $\Phi$ (needed to forward-model geometry effects):
\begin{equation} \label{eq:Phi-estimator}
    \Phi^b_\ell = \frac{1}{\overline{V(nw)^2}}\frac{2\ell+1}{V_b}\sum_{i}\sum_{j,\, i \neq j} w_iw_j\mathcal{L}_\ell(\hat{\mat r}_i\cdot\hat{\mat r}_j)\Theta^b(|\mat r_i-\mat r_j|) \, ,
\end{equation}
restricting to radial bin $b$ and Legendre multipole $\ell$. 
The measurement of the multipoles of the power spectrum and corresponding window functions for a data set with $\sim 10^8$ pixels (similar to the data set used in this work) takes $\sim 5$ hours on the Perlmutter computer at NERSC (using one AMD Milan CPU).

\section{Theory Power Spectrum Modeling} \label{sec:theory_pk}
One of the key advantages of the \Lya forest is that it probes the Universe at redshifts $2 \leq z \leq 5$, \ie, with many more linear modes than seen in galaxy surveys. For our theory modeling, we use the same procedure to the one outlined in \citetalias{dMdB:2020} and briefly summarize it below.\footnote{Note that we do not include the $G(\mat k)$ in Eq.~(27) of \citetalias{dMdB:2020} since we do not grid the data for our power spectrum computation.} The real-space quasi-linear power spectrum\footnote{The linear input power spectrum is computed using the Boltzmann solver \texttt{CAMB} (\url{https://camb.info/}) as part of publicly available \texttt{vega} package (\url{https://github.com/andreicuceu/vega/}).}, $P_{\rm QL}(k,\mu,z)$, is modulated by a simulation-based non-linear fitting function, $F_{\rm{NL}}$, and connected to the redshift-space flux power through the well-known Kaiser formula \citep{Kaiser1987}
\begin{equation} \label{eq:Kaiser}
    P_F(k,\mu, z) = b_{F}(z)^2\left(1+\beta_F \mu^2\right)^2 F_{\rm{NL}}(k,\mu) P_{\rm QL}(k,z) \ ,
\end{equation}
where $b_F$ denotes the redshift-dependent linear bias parameter, $\beta_F$ is the redshift-space distortion (RSD) parameter (which is assumed to be redshift-independent) and $\mu$ is the angle of $\kvec = \{\kpar, \kvperp\}$ to the line-of-sight, $\mu \equiv \kpar/k$. The redshift-evolution of the \Lya forest bias parameter enters the weights, $w$, which are taken from Eq.~7 in \citetalias{dMdB:2020}, and enters the power spectrum measurement through Eqs.~\eqref{eq:DD_pk} and~\eqref{eq:Phi-estimator}. Following \cite{McDonald06}; \citetalias{dMdB:2020}; \cite{2023JCAP...11..045G}, the product of \Lya forest bias and growth factor is assumed to have a redshift dependence described by $(1+z)^{\gamma_{\text{\Lya}}-1}$ where $\gamma_{\text{\Lya}}=2.9$ is the redshift evolution parameter of the \Lya forest bias.
Over sufficiently large scales the \Lya forest fluctuations are, to linear order, given by $\td_F = b_{\td,F}\td+b_{\eta,F}\eta$ where $\td$ are mass fluctuations and $\eta=-(\dd v_p/\dd x_p)/aH$ is the gradient of the peculiar velocity $v_p$ over the comoving line-of-sight coordinate $x_p$ with the scale factor $a$ and the Hubble constant $H$. As such, the bias factors are the partial derivatives with respect to $\td$ and $\eta$ \citep[see, e.g.,~][]{Arinyo:2015}. The redshift space distortion parameter for \Lya forest surveys is 
\beq \beta_F = \frac{f b_{\eta,F}}{b_{\delta,F}} \ ,
\eeq
for our fiducial cosmology we obtain $f = 0.9704$. We can estimate $b_\eta$ and $b_\delta$ from simulations, by either fitting the (i) auto-power spectrum using the Kaiser formula \citep[e.g.,][]{Arinyo:2015}; (ii) cross-power of \Lya and matter \citep[e.g.,][]{Givans:2022}; or (iii) from separate Universe simulations \citep[e.g.,][]{Cieplak:2016}.  For the \Lya forest this results in an RSD parameter greater than unity, yielding a quadrupole that is larger than the monopole (in contrast to the case for galaxies). To account for the broadening in $\kpar$ stemming from high column density (HCD) systems \citep[see, e.g.,~][and references therein]{Rogers:2018}, the \Lya bias parameters $(b,\beta)$ are remapped to
\begin{align}
    b'_F &= b_F + b_{\rm HCD}F_{\rm HCD}(\kpar) \ , \\
    b'_F\beta'_F &= b_F\beta_F + b_{\rm HCD}\beta_{\rm HCD}F_{\rm HCD}(\kpar) \ ,
\end{align}
where $F_{\rm HCD}(\kpar)=\exp(-L_{\rm HCD}\kpar)$ is fitted to hydrodynamical simulations and $L_{\rm HCD}$ is the typical scale for unmasked HCDs, set to $10 \hinvMpc$ in \citetalias{dMdB:2020}. 

The non-linear (NL) correction to the power spectrum is obtained from fits to hydrodynamical simulations \citep{Arinyo:2015}
\begin{equation} \label{eq:Fnl}
    F_{\rm{NL}}(k,\mu) = \exp \left( q_1\Delta^2(k) \left[ 1-\left(\frac{k}{k_{\nu}}\right)^{\alpha_v}\mu^{b_{\nu}} \right] - \left(\frac{k}{k_{p}}\right)^2\right)\ ,
\end{equation}
with the usual dimensionless power spectrum $\Delta^2(k) \equiv k^3P_{\rm lin}(k)/(2\pi^2)$. The parameter $q_1$ is a dimensionless parameter and controls the importance of the non-linear enhancement.\footnote{The parameter $q_1$ determines the impact of non-linearities, and is related to the parameter $k_{\rm NL}$ in \cite{McDonald:2001}.} While $q_1$ is only (very) weakly dependent on redshift, it is, together with $\alpha_v$ and $k_{\nu}^{\alpha_v}$, inversely proportional to the amplitude of the linear power spectrum. The remaining parameters model the Jeans smoothing and line-of-sight broadening stemming from non-linear peculiar velocities and thermal broadening \citep[see ][]{McDonald:2001, Arinyo:2015}. We set the parameters to $q_1=0.8558$, $k_{\nu}=1.11454 \hinvMpc$, $\alpha_{\nu}=0.5378$, $b_{\nu}=1.607$ and $k_p=19.47 \hinvMpc$, following \citetalias{dMdB:2020}. 

The quasi-linear power spectrum is given by 
\begin{equation}
    P_{\rm QL}(k,z) = P_{\rm sm}(k,z)+\exp{\left[-\frac{\kpar^2\Sigma_{\parallel}^2 + \kperp^2\Sigma_{\perp}^2}{2}\right]}P_{\rm peak} (k,z) \ ,
\end{equation}
which decomposes the power spectrum into a smooth (no BAO feature) and a peak (isolating the BAO feature) component. This description includes a correction for the non-linear broadening of the BAO peak denoted by $\Sigma_{\parallel}/\Sigma_{\perp} = 1+f$ where $f$ is the (linear) growth rate \citep[see, e.g.,~][]{Eisenstein2007}. Analogously to \citetalias{dMdB:2020}, we set the smoothing parameters to $\Sigma_{\perp}=3.26 \hinvMpc$ and $\Sigma_{\perp}=6.42\hinvMpc$ for a growth rate of $f\approx 0.97$.

Contaminants in the \Lya forest, such as absorption by metals and correlations due to the sky-subtraction procedure of the eBOSS data pipeline, are modeled as additive terms in the \Lya auto-correlation function \citepalias[see Sec. 4 in][for a detailed discussion]{dMdB:2020}. The computation of comoving separations is based on the (erroneous) assumption that all absorption stems solely from the \hi \Lya transition. Transitions from other elements (\eg silicon and carbon often denoted as metals) contribute to the measured \Lya forest fluctuations as well. Thus, the measured absorption at a given wavelength is a mixture of absorption at different redshifts (from different transitions). For each pair of possible transitions the offset between the true and assumed redshift is computed. The resulting re-mapping matrix between the true and assumed comoving separation is called the metal matrix, see, \eg,~\cite{Blomqvist_2018}, \citetalias{dMdB:2020}, relating the measured to the `true' summary statistic in configuration space, \eg~the correlation function.

To estimate the window-convolved power spectra, we will require the correlation function multipoles: these are defined as an angular integral over the full correlation function $\xi$,
\begin{equation} \label{eq:RSD}
    \xi_{\ell}(r) \equiv \frac{2\ell +1}{2} \int \displaylimits_{\mu =-1}\displaylimits^{+1}\mathrm{d}\mu\, \xi(r,\mu) \mathcal{L}_{\ell}(\mu)\ .
\end{equation}
As described below, these will then be used to compute the mode mixing introduced by the window matrix in Eq.~\eqref{eq:PWell}, allowing accurate comparison of theory and data. 

\subsection{Forward modeling the window function} \label{Sec:window_func}
In the present analysis, we forward model the effect of the survey geometry, captured by the configuration-space window matrix $\Phi(\vr)\equiv DD(\vr)/\xi(\vr)$, on the theoretically expected power spectrum instead of removing the window function from the data. This factor is independent of the \Lya fluctuations, $\delta_F$, and can be explicitly computed by counting pairs of \Lya pixels, as in Eq.~\eqref{eq:Phi-estimator}. This is analogous to the approach used in galaxy surveys e.g.,~\cite{Castorina2018, Beutler:2016, 2021JCAP...11..031B}, whence one writes
\begin{equation} 
    \Phi(\mat r) \equiv \frac{RR(\mat r)}{RR_{\rm model}(\mat r)} = \frac{RR(\mat r)}{\overline{V(nw)^2}} \ ,
\end{equation}
for (weighted) background density $R(\mat x) = n(\mat x)w(\mat x)$, and an ideal bin volume $RR_{\rm model}(\vr) = \frac{4\pi}{3} (r_{\rm max}^3 - r_{\rm min}^3)$, given some set of (thin) bins in $r_{ij}$. In the galaxy case, this is estimated using catalogs of random particles; for us, the procedure simply involves counting the unweighted \Lya pixels.\footnote{In full, the \Lya case can be imagined as a set of (correlated) pencil-beam surveys, whose sampling is known precisely.}

Expressed in terms of (even) Legendre multipoles, the pair-count estimator of Eq.~\eqref{eq:DD_pk} measures the power spectrum convolved with the functions
\begin{equation}
\Gamma_{\ell}^2(r) \equiv \Phi_{\ell}(r)W(r;R_0) \ ,
\label{eq:Well}
\end{equation}
where $W(r;R_0)$ is the pair-truncation function given in Eq.~\eqref{eq:wofr}. The convolved power spectrum multipoles can be obtained by first considering the distortion induced to the 2PCF (see, e.g.,~\cite{Beutler:2016}):
\begin{align} 
	\tilde{\xi}_{0}(r) = \xi_{0}\Gamma_{0}^2  &+ \frac{1}{5}\xi_2\Gamma^2_2 + \frac{1}{9}\xi_4\Gamma^2_4 + \dots
	\label{eq:conv1}\\
	\begin{split}
		\tilde{\xi}_{2}(r) = \xi_{0} \Gamma_{2}^2  &+ \xi_2\left[\Gamma^2_0 + \frac{2}{7}\Gamma^2_2 + \frac{2}{7}\Gamma^2_4\right]\\
		&+\xi_4\left[\frac{2}{7}\Gamma^2_2 + \frac{100}{693}\Gamma^2_4 + \frac{25}{143}\Gamma^2_6\right] + \dots
		\label{eq:conv2}
	\end{split}\\
	\begin{split}
		\tilde{\xi}_{4}(r) = \xi_{0} \Gamma_{4}^2  &+ \xi_2\left[\frac{18}{35}\Gamma^2_2 + \frac{20}{77}\Gamma^2_4 + \frac{45}{143}\Gamma^2_6\right]\\
		&+\xi_4\left[\Gamma^2_0 + \frac{20}{77}\Gamma^2_2 + \frac{162}{1001}\Gamma^2_4 \right.\\
		&\left. \,\,\,\,\,\,\,\,\,\, \qquad + \frac{20}{143}\Gamma^2_6 + \frac{490}{2431}\Gamma^2_8 \right]+ \dots
		\label{eq:conv3}
	\end{split}
\end{align}
From Eq.~\eqref{eq: pk-conv}, the power spectra can then be expressed as a Hankel transform:
\begin{equation} 
\tilde{P}^a_{\ell} = 4\pi i^\ell \int_0^{R_0} \diff r\,r^2\tilde{\xi}_{\ell}(r) K^a_\ell(r) \ ,
\label{eq:PWell}
\end{equation}
where $K_\ell^a(r)$ is the usual spherical Bessel function $j_\ell(kr)$ integrated over a $k$-bin to reduce oscillations:
\beq 
    K_\ell^a(r) = \frac{\int_{k_{\rm min}^a}^{k_{\rm max}^a} k^2\diff k\,j_\ell(kr)}{\int_{k_{\rm min}^a}^{k_{\rm max}^a} k^2 \diff k} \ ,\eeq
as in Eq.~\eqref{eq: k-binning}.\footnote{These integrals of the spherical Bessel function are analytic (see, e.g.,~\url{https://dlmf.nist.gov/}).} For a thin $k$-bin centered at $k_a$, $K_\ell^a(r)\approx j_\ell(k_ar)$.
In this work, we truncate our expansion of the window-convolved correlation function, $\tilde{\xi_{\ell}}$, at $\ell_{\rm max}=4$ for the 2PCF. Note that we include \textit{all} corresponding window matrix multipoles for each multipole in $\tilde{\xi_{\ell}}$.\footnote{We tested that including 2PCF contributions up to $\ell_{\rm max}=6$ with all corresponding window matrix multipoles up to $\ell_{\rm max}=10$ did not affect our results.} In Appendix \ref{sec:appendix_window_func}, we discuss the contribution of each correlation function multipole to the final result.

\section{Lyman-a forest data from eBOSS DR16} \label{sec:data}
In the present analysis, we compute the 3D power spectrum from \Lya forest spectra from the 16th data release \citep[DR16Q;][]{eBOSS_DR16Q} of the completed extended Baryonic Oscillation Spectroscopic Survey \citep[eBOSS;][]{Dawson:2016} of the fourth generation of the Sloan Digital Sky Survey \citep[SDSS-IV;][]{York2000}. The primary scientific goal of eBOSS was to constrain cosmological parameters using BAO and redshift space distortions \citep[see, e.g.,][]{Alam:2021, dMdB:2020}. Briefly summarized, the DR16 data consists of the complete 5-year BOSS and 5-year eBOSS survey. The \Lya spectra were observed with a double spectrograph mounted on the 2.5m Apache Point telescope to map 10,000 deg$^2$ of the sky. The observations are conducted in the observed wavelength range of $3600 < \lamobs < 10,000\,$\AA\ with a spectral resolution of $R\sim 1500-2500$ \citep{Dawson:2016}. The quasar selection and algorithms of the pipeline are explained in detail in \cite{2015ApJS..221...27M} as well as technical details related to the survey itself in \cite{Dawson:2016}. 

\subsection{Data selection}
The full sample consists of $205,012$ quasar spectra with absorber redshifts in the range $1.96\leq z\leq 3.93$. The quasar sample is split into two disjoint regions on the sky; the northern (NGC) and southern (SGC) galactic caps, with $147,392$ and $57,620$ sight lines respectively.  The number of spectral pixels are $34.3 \cdot 10^6$ for the entire data set with an average signal-to-noise ratio of $2.56$ per pixel calculated in the \Lya forest. The forest used for cosmological analysis is defined to be $1040 \leq \lamrf \leq 1200\,$\AA. For ease of comparison to \citetalias{dMdB:2020}, we use the same set of flux decrements, $\delta_F$, introduced in the following section. We note that BOSS/eBOSS observed spectra in spectral pixels of with $\Delta \log_{10}(\lambda)\sim 10^{-4}$ which have been rebinned for the purpose of the final analysis onto a grid of $\Delta \log_{10}(\lambda)\sim 3\times 10^{-4}$.  Damped \Lya absorbers (DLAs), defined as regions where the transmission is reduced by more than 20\% in the flux decrements, are masked out. A Voigt profile is fitted to each DLA to correct for the absorption in the wings \citep{2012A&A...547L...1N}. Quasars with broad absorption lines (BALs) have been removed for the present analyses \citep[see ][for more details]{eBOSS_DR16Q}. Both are the main contaminants affecting the data selection.

\subsection{Continuum fitting} \label{sec:cont_fit}
To measure clustering statistics such as the P3D, we use the flux decrement given in Eq.~\eqref{eq:delta_F}. Extracting the unabsorbed flux, \ie, the quasar continuum, is a daunting task in general. In the present paper, we use the publicly available continuum-fitted spectra from eBOSS DR16\footnote{Publicly available at \url{https://data.sdss.org/sas/dr16/}.}. In the following, we briefly outline the continuum fitting procedure from \citetalias{dMdB:2020}: to each \Lya forest a slope and an amplitude is fitted to the stacked spectra in the quasar sample using the \texttt{picca} package.\footnote{Publicly available at \url{https://github.com/igmhub/picca}.} The product in the denominator of Eq.~\eqref{eq:delta_F} assumes a common continuum for all quasars in restframe, $\overline{q}(\lamrf)$, corrected by a first order polynomial for each quasar $q$ in $\Lambda \equiv \log{\lambda}$ which accounts for the diversity of the quasars
\begin{equation}
    \overline{F}(\lambda)q(\lambda) = \overline{q}(\lambda)\left(a_q+b_q \frac{\Lambda - \Lambda_{\rm min}}{\Lambda_{\rm max} - \Lambda_{\rm min}}\right) , 
\end{equation}
which results in a biased mean and spectral slope of each forest flux decrement $\Hat{\td}(\lambda) = F(\lambda)/\left[(a_q+b_q\Lambda(\lambda))\overline{q}(\lambda)\right] -1$. The fluctuations are then given by 
\begin{equation} \label{eq:delta_distort}
    \tilde{\td}(\lambda) = \Hat{\td}(\lambda) - \frac{1}{W_q}\sum_{\lambda'}w(\lambda')\Hat{\td}(\lambda')\left[1+\frac{\Lambda(\lambda)\Lambda(\lambda')}{\overline{\Lambda^2(\lambda)}}\right]\ ,
\end{equation}
with $W_q = \sum_{\lambda'}w(\lambda')$ and $\Lambda(\lambda) = \log{\lambda}-\overline{\log{\lambda}}$. We use this quantity to measure the power spectrum. Estimating the continuum directly from the forest distorts the field, requiring correction at the level of the flux decrement (see, e.g.,~\citetalias{dMdB:2020}). The continuum fitting and projection to center the mean flux decrement at zero suppresses modes along the line-of-sight in the forest. A good approximation is to treat these distortions as linear and model them in the theoretically expected correlation function through a distortion matrix \citepalias[DM;][]{dMdB:2020} 
\begin{equation} \label{eq:DM}
    \xi^{\rm DM}_{\rm A} = \sum_{\rm A'} D_{\rm AA'} \xi_{\rm A'} \ ,
\end{equation}
where $A$ and $A'$ denote two bins of the correlation function
\begin{equation}
    D_{\rm AA'} = \frac{1}{\sum_{\lambda}w(\lambda)} \sum_{(i,j)\in A} w_iw_j \sum_{(i',j')\in A'}\eta_{ii'} \eta_{jj'} \ , 
\end{equation}
with \beq \eta_{ii'}\equiv\td^{\rm K}_{ii'}-\frac{w_j}{\sum_k w_k} - \frac{w_j\Lambda_i\Lambda_j}{\sum_k w_k\Lambda_k^2} \ , \eeq for forest pixel pairs $(i,j)$ and $(i',j')$ in sight lines $A$ and $A'$ . Whilst this approach marginalizes out large-scale modes, it lends itself well to measure the small-scale continuum for the one-dimensional power spectrum. In the current implementation we apply the distortion matrix to the theory correlation function in $\{\rpar,\rvperp\}-$space.\footnote{Analogous to \citetalias{dMdB:2020} we measure the distortion matrix in bins of $4\hinvMpc$ in the range $0 \leq \{\rpar, \rperp\} \leq 200 \hinvMpc$.} (We denote the theory correlation function to be `distorted' after applying the distortion matrix.) The distorted multipoles are then used to compute the window convolved theory power spectrum in Eq.~\eqref{eq:PWell}. See Appendix \ref{sec:DM} for the effect of the distortion matrix on theory power spectra and correlation function multipoles. 

\section{Consistency Tests on Simulations} \label{sec:simulations}
Before presenting the main results of this paper, we first test our pipeline by applying the power spectrum estimator, presented in Sec.~\ref{sec:methods}, to idealized Gaussian realizations, with results displayed in Sec.~\ref{sec:GRF}. In Sec.~\ref{sec:lyacolore} we apply our pipeline to realistic eBOSS DR16 mocks including contaminants (such as high column density absorbers and metals). In particular, we discuss the effect of continuum fitting and the extraction of the unabsorbed flux, on the resulting power spectrum measurements.

\subsection{Gaussian random field simulations} \label{sec:GRF}
The first series of tests of the power spectrum estimator presented in Sec.~\ref{sec:methods} is on anisotropic Gaussian random fields (GRFs), generated with a known linear matter power spectrum $P_{\rm lin}(k)$. The aim is to investigate biases as well as the range of validity and robustness of the estimator to non-trivial survey geometries with a known input matter power spectrum.\footnote{In the following section, we will discuss a series of tests on more realistic mocks.} The GRFs are realized in a periodic box of size $L = 1380 \hinvMpc$ with $N=512$ cells at redshift $z=2.4$ with a $\Lambda$CDM linear power spectrum as theory input \citep{Aghanim:2018eyx}. The fundamental mode is $k_F = 2\pi / L = 0.0045\, \hMpcinv$ with the Nyquist frequency being $k_{\rm Ny} = 1.17\, \hMpcinv$. 

We sample $N_{\rm s}=900$ sight lines from the box, corresponding to a (very sparse) quasar density of $\sim$\,$2-3~\mathrm{qso}/\mathrm{deg}^2$, and $460,800$ \Lya pixels.\footnote{We performed the same test using a DESI-like quasar density of $\sim$\,$30~\mathrm{qso}/\mathrm{deg}^2$ and recover the input power spectrum to percent-level precision over the same $k$-range.} For the \Lya forest fluctuations, $\delta$, we use the pixel value, \ie, the modes from the GRF, and model the selection function using weights of unity. For the power spectrum measurement, we do not include pixels in the same skewer, as in the main \Lya analysis. Additionally, we apply a non-trivial selection function, \ie, with edges and sparse sampling transverse to the line-of-sight to measure the effect of the window function on the measured power spectrum. To this end, we remove the periodicity of the box and add a 10\% root-mean-squared error, relative to the mean, Gaussian, white (uncorrelated), noise to each of the pixels in the box, corresponding to somewhat realistic \Lya noise levels for continuum fitting and pipeline noise. 

\begin{figure}
    \centering
    \includegraphics[width=1\linewidth]{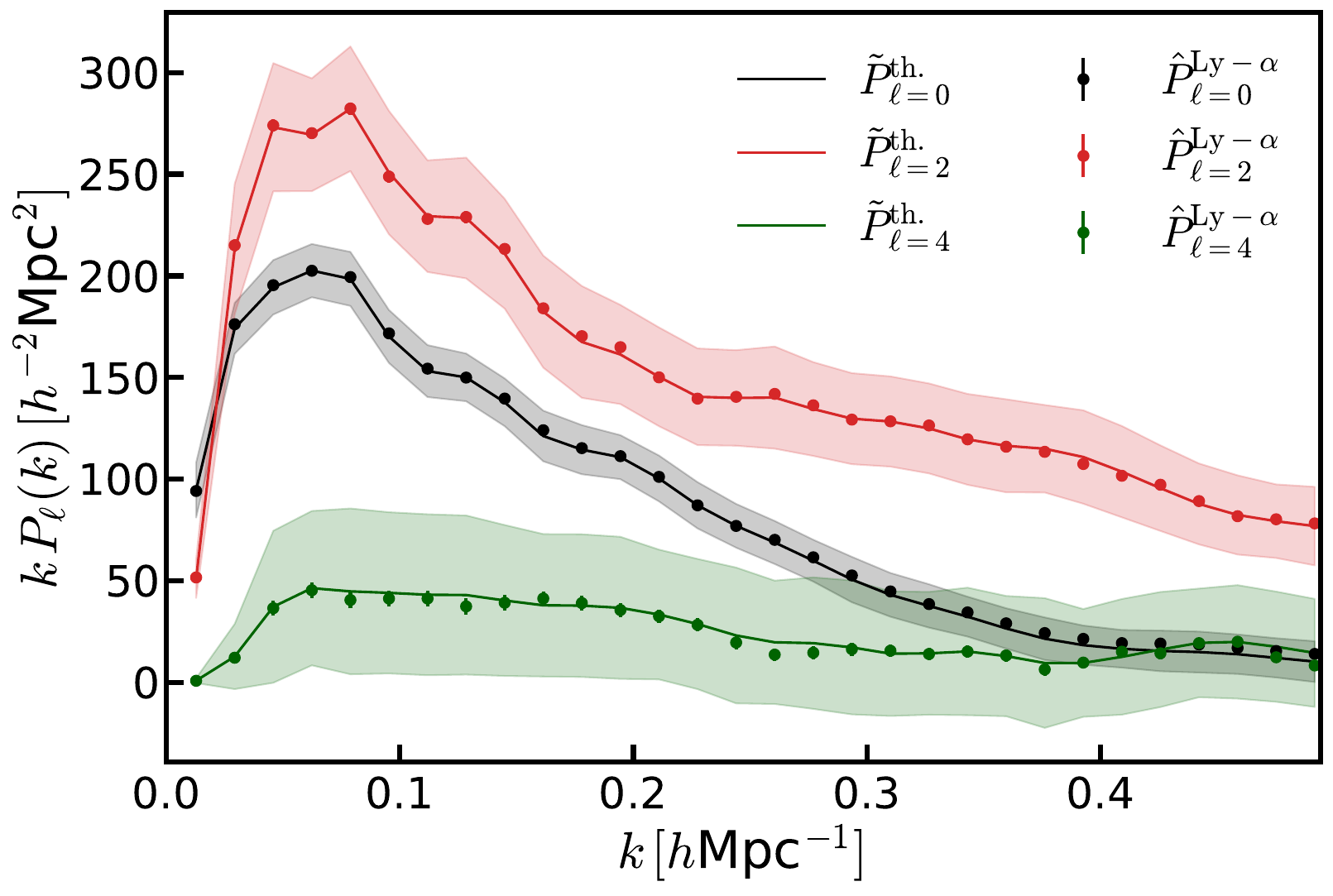}
    \vspace{-0.1in}
    \caption{Consistency test of the power spectrum estimator, comparing the window-convolved linear power spectrum at $z=2.4$ (monopole in black, quadrupole in red, hexadecapole in green, respectively) to the mean power spectrum of $N=100$ Gaussian random field simulations (dots), computed using pair counts truncated $R_0=200\hinvMpc$. The shaded regions give the square root of the diagonal of the covariance matrix and illustrate the variance between the $N=100$ realizations; this is also shown by the error-bars, which include the factor of $1/\sqrt{N}=1/10$. 
    The 2PCF contributions are computed up to $\ell_{\rm max}=4$ with all corresponding window matrix multipoles, as described in Sec.~\ref{sec:theory_pk}.}  
    \label{fig:p3d_grf}
    \vspace{-0.1in}
\end{figure}

We measure the monopole, quadrupole and hexadecapole power spectra from the GRFs and compute the 2PCF contributions up to $\ell_{\rm max}=4$ with all corresponding window matrix multipoles. We use $30$ equidistant $k$-bins in the range $(0.01 \leq k \leq 0.50) \hMpcinv$ and $N_r=1,000$ linearly spaced bins in the range $(0 \leq r \leq R_0=200) \hinvMpc$ for the window matrix, respectively.\footnote{We tested that our results are robust to the choice of $N_r$ and the $\ell_{\rm max}$ of the 2PCF (and the corresponding window matrix multipole).} We compare the measured power spectrum multipoles to the theory input power spectrum in Fig.~\ref{fig:p3d_grf} and recover the input spectrum well within the $1\sigma$ error bars up to $k \simlt 0.5 \hMpcinv$ with the measured anisotropic power spectra. Note that the quadrupole is larger than the monopole due to the large RSD parameter $\beta_F = 1.5$, typical for \Lya surveys. Note that the pair separation window $W(r;R_0)$ suppresses power on scales larger than the ones defined by $k = 2\pi/R_0$ which is at $k = 0.031 \hMpcinv$ for our chosen value of $R_0$. Thus, we discard $k$-values below the cutoff in our analysis pipeline. In Appendix \ref{sec:appendix_window_func}, we discuss the effect of the number of skewers on the multipoles of the window matrix itself as well as their convolution with the window convolved theory power spectra.\footnote{Note that we use the terms `window function' and `window matrix' interchangeably.} 

\begin{figure}
    \centering
    \includegraphics[width=\linewidth]{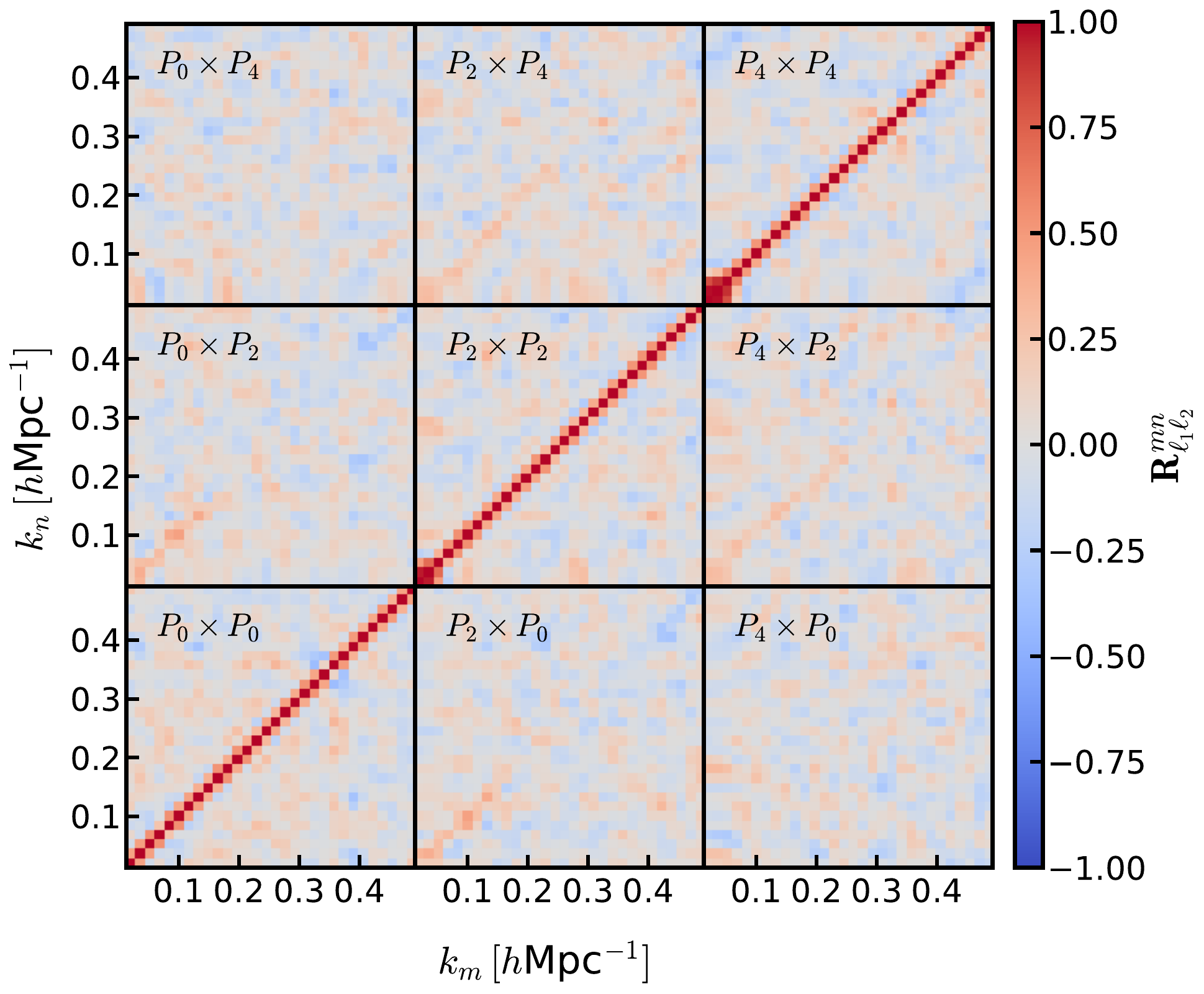}
    \vspace{-0.1in}
    \caption{Correlation matrix of the power spectrum multipoles $P_{\ell_1}\times P_{\ell_2}$ measured from $N=100$ Gaussian random field realizations, as in Fig.\,\ref{fig:p3d_grf}. The correlation matrix is defined in Eq.~\eqref{eq:corr_mat}; red denotes fully correlated and blue fully anti-correlated power spectrum bins.}
    \label{fig:corr_mat}
    \vspace{-0.1in}
\end{figure}

The error bars shown in Fig.~\ref{fig:p3d_grf} capture the variation between differing GRF realizations and are obtained from the diagonal of the covariance matrix computed from $N=100$ realizations of the GRFs. For Legendre multipoles $\ell_1, \ell_2$ of the anisotropic power in $k$-bins $m, n$, this is given by
\begin{equation} \label{eq:cov_mat}
    \mat{C}_{\ell_1\ell_2}^{mn} = \frac{1}{N-1} \sum_{i=1}^N \left(\hat{P}^{m}_{\ell_1,i}-\Bar{P}^{m}_{\ell_1}\right)\left(\hat{P}^{n}_{\ell_2,i}-\Bar{P}^{n}_{\ell_2}\right) ,
\end{equation}
with the mean of the power spectra given by
\begin{equation} \label{eq:pk_mean}
\Bar{P}^{m}_{\ell} = \frac{1}{N}\sum_{i=1}^N\hat{P}^{m}_{\ell,i}\ ,
\end{equation}
where $i$ indexes the $N$ simulations. The correlation matrix is defined as
\begin{equation} \label{eq:corr_mat}
    \mat{R}^{mn}_{\ell_1\ell_2} \equiv \frac{\mat{C}_{\ell_1\ell_2}^{mn}}{\sqrt{\mat{C}_{\ell_1\ell_1}^{mm}\mat{C}_{\ell_2\ell_2}^{nn}}} \ ;
\end{equation}
by construction, this is unity along the diagonal. We show all combinations of the correlation matrix moments in Fig.~\ref{fig:corr_mat}: the first column displays the correlation between $k$-bins in the monopole with itself, the quadruple and the hexadecapole (from bottom to top, respectively). Analogously, the second column displays the correlations of the quadrupole with the same power spectrum multipoles and the third column shows the correlations of the hexadecapole with said moments. The anisotropies sourced by the imposed survey geometry and intrinsic correlations from the generated anisotropies are (weakly) visible as correlations along the diagonal between the monopole and quadrupole (and the quadrupole with the hexadecapole, respectively.). We find negligible correlation between the monopole and hexadecapole (the latter being mostly dominated by noise). For the $\ell_1=\ell_2$ correlations, we observe a correlation length of up to two $k$-bins, resulting in an approximately bi-diagonal correlation matrix. Note that for the quadrupole and the hexadecapole, the lowest three $k$-bins are highly correlated. This can be removed by increasing $R_0$ in the pair separation function, though we caution that computation time scales as $\mathcal{O}(N_{\rm pix}^2R_0^3)$. 

Whilst the above tests on GRFs give us confidence that the P3D estimator recovers the input power spectrum to high precision in the presence of survey geometries typical for \Lya data analysis, we stress that these simplistic simulations ignore crucial systematic and instrumental effects as well as details of the data reduction pipeline present in real data, e.g.,~the co-addition of individual spectra, per spectral pixel noise estimates, sky residuals, continuum fitting, metal contamination, damped \Lya absorbers and broad absorption lines. Whilst resolution mostly affects high-$k$ modes, the key systematic arising along the line-of-sight, \ie~at large scales, is the uncertainty in the continuum measurement.\footnote{See \cite{Font-Ribera:2018, Horowitz:2024} for approaches on how to marginalize over continuum uncertainties in the power spectrum computation.} In the following section, we address (some of) these issues using more realistic \Lya forest synthetic data. 

\subsection{Synthetic \Lya spectra: \Lyacolore simulations}\label{sec:lyacolore} 
In this section, we present tests of our estimator on realistic \Lya forest simulations (for a more complete discussion of the \Lyacolore mocks, see \citet{Farr20} and \citetalias{dMdB:2020}). In the following we will briefly summarize the key steps to generate these simulations: Each realization is based on a GRF of length $10\hinvGpc$ and $4096^3$ particles yielding a resolution of $\sim 2.4\hinvMpc$ generated using \texttt{CoLoRe}'s log-normal density model \citep{2022JCAP...05..002R}.\footnote{Publicly available at \url{https://github.com/damonge/CoLoRe}.} The input cosmology is based on the best fit $\Lambda$CDM-parameters obtained by the \Planck satellite \citep[see column 1 of table 3 in ][]{2016A&A...594A..13P}. The Gaussian field is subsequently transformed to a log-normal density field which is Poisson sampled assuming a quasar density of 59 quasars per $\deg^2$ and a functional form for the quasar bias redshift evolution (see, e.g.,~\citetalias{dMdB:2020}). The line-of-sight skewers are computed by interpolating from the GRF onto the `observed' pixel positions and the radial velocity field to the center of the box. The obtained skewers are then post-processed using the \Lyacolore package \citep{Farr20, 2024arXiv240100303H}.\footnote{Publicly available at \url{https://github.com/igmhub/LyaCoLoRe}.} The sparse sampling of the \Lya forest transverse to the line-of-sight (also denoted by $\td$-sampling) results in power on small scales contributing to the error on large scales \citep[see, e.g.,~][]{McDonald:2007} which is quantified by the one-dimensional power spectrum (P1D). Thus, additional power is added to each skewer by sampling from a Gaussian with variance set by the P1D. From the log-normal transformation of the final skewer the optical depth field, $\tau$, is computed using the fluctuating Gunn-Peterson approximation \citep[FGPA;][]{1998ApJ...495...44C}. Redshift space distortions are obtained by convolving the $\tau$ field with the peculiar velocity field. The observed flux is then related to $\tau$ via $F=\exp{(-\tau)}$. 

The \Lyacolore mocks mimic noise and instrumental systematics present in the eBOSS data.\footnote{Note that while this set of synthetic spectra have been tuned to BAO-scale analyses and have also been tested in the context of full-shape analyses of the \Lya forest with the 2PCF \citep[see, e.g.,~][]{Cuceu:2023}, these have not been explicitly developed for the purpose of P3D analyses.} In particular, the effect of continuum fitting, instrumental noise, high column density (HCD) absorbers\footnote{Following \citetalias{dMdB:2020}, we denote systems with neutral hydrogen exceeding $10^{17.2}\, \mathrm{atoms}\,\mathrm{cm}^{-2}$ as HCDs.}, \Lya absorption and metals. Additionally, random redshift errors for the spectra are included by drawing from a Gaussian with mean zero and $\sigma_z = 400\, \mathrm{km/s}$. The eBOSS analysis pipeline and spectral resolution are simulated using the \texttt{specsim} package  \citep{david_kirkby_2021_4566008}.\footnote{Publicly available at \url{https://github.com/desihub/specsim}.} The co-addition of the resulting skewers with instrumental noise and a realistic quasar continuum is done using the \texttt{desisim} package.\footnote{Publicly available at \url{https://github.com/desihub/desisim}.} 

In the following, we present tests of our power spectrum estimator on $N$ realizations of mocks with increasing levels of systematics. We compare four different set of mocks (and use a similar nomenclature as the one adopted in table 4 of \citetalias{dMdB:2020}): 
\begin{enumerate}
    \item \Lya raw mocks (eboss-raw): We estimate the P3D directly from the simulated \Lya forest fluctuations. This analysis is similar to the case of the GRFs but using the eBOSS \Lya survey geometry with the corresponding effective volume. We use $N=10$ realizations.
    \item distorted \Lya raw mocks (eboss-raw-dist): We include a variation of the `raw' mocks of which we remove the mean of each skewer and denote them by `raw-dist'. We use $N=10$ realizations.
    \item +continuum+noise (eboss-0.0): We add the effect of measuring quasar continua of the spectra and include instrumental noise. The performed continuum fitting, introduced in Sec.~\ref{sec:cont_fit}, requires forward modeling the suppression of modes along the line-of-sight through a so-called distortion matrix given in Eq.~\eqref{eq:DM}. We use $N=10$ realizations.
    \item + metals + HCDs + $\sigma_{\nu}$ (eboss-0.2): We estimate P3D from realistic \Lya mock spectra including the effects stemming from continuum fitting and instrumental noise (from step ii) and adding metals, high column density absorbers and random redshift errors. Analogous to the analysis on real data, HCDs are masked based on the cut $\log{N_{\text{\hi}}}>20.3\, \mathrm{cm}^{-2}$. We use $N=100$ realizations.
\end{enumerate}

\begin{figure}
    \centering
    \includegraphics[width=1\linewidth]{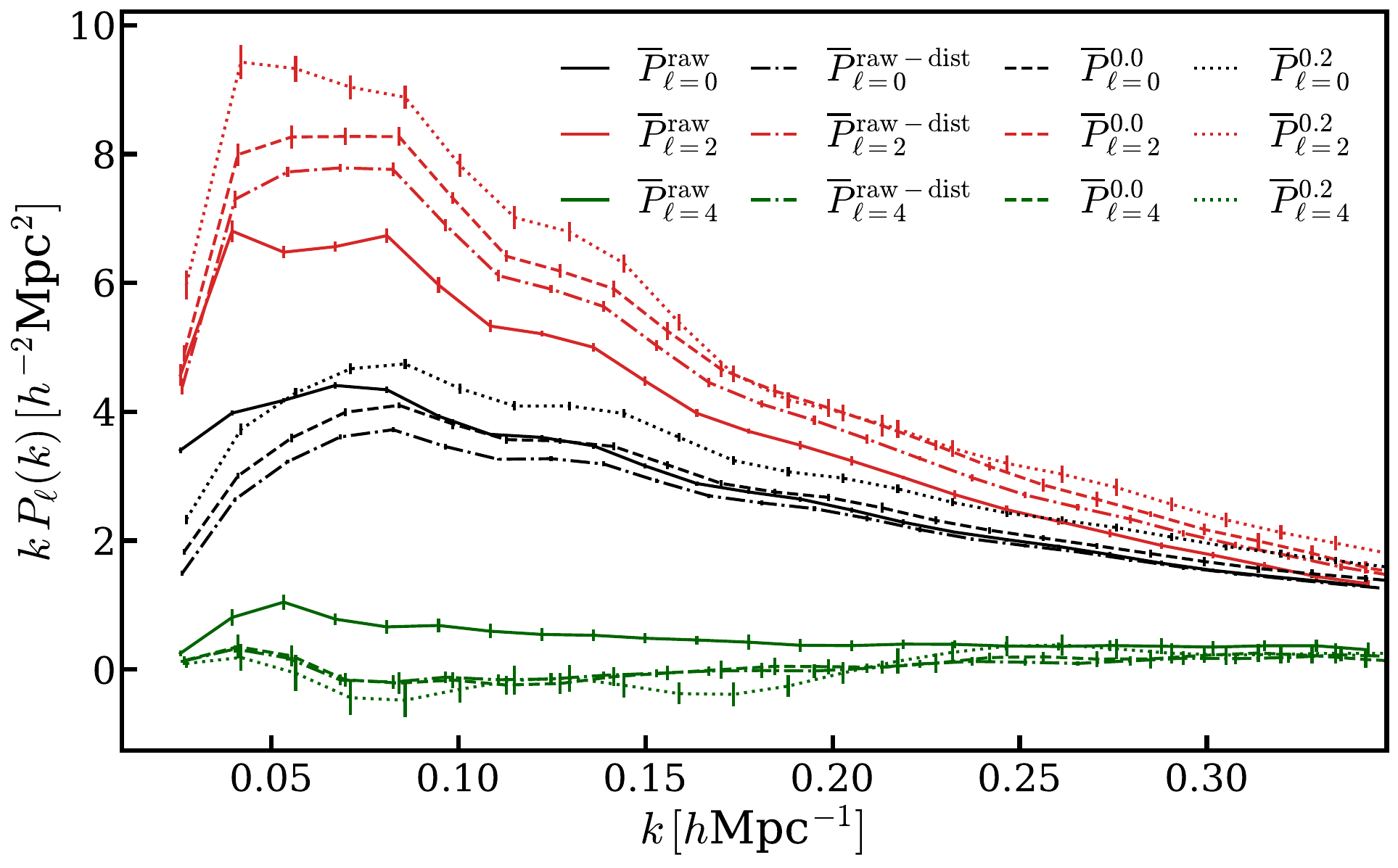}
    \vspace{-0.1in}
    \caption{Comparison of \Lya power spectra measured from \Lyacolore mocks for four different configurations: `raw' are the raw \Lya forest fluctuations (solid lines); `raw-dist' are the distorted fluctuations where the mean and the slope along each sight line have been removed (dashed-dotted lines);  `0.0' include continua and instrumental noise (dashed lines); `0.2' include metals, HCDs and random redshift errors (dotted lines). The colored lines (black for the monopole, red for the quadrupole and green for the hexadecapole, respectively) are the mean power spectra and the error bars are the variance between the different realizations.}
    \label{fig:mocks_P3D}
    \vspace{-0.1in}
\end{figure}

In Fig.~\ref{fig:mocks_P3D} we compare the power spectra of the \Lyacolore mocks for the four different mock configurations, alongside their errors. As in the main analysis, we measure the power spectra up to $\ell_{\rm max}=4$ and compute the window function multipoles up to $\ell_{\rm max}=8$. Therefore, we use $24$ equidistant $k$-bins in the range $(0.02 \leq k \leq 0.35) \hMpcinv$ and $N_r=1,000$ linearly spaced bins in the range $(0 \leq r \leq R_0=200) \hinvMpc$ for the window matrix, respectively. We treat the mean power spectra measured from the `raw' \Lya forest fluctuations mocks (solid colored lines in the plot), as ground truth to assess the effect of contaminants on the power spectra.\footnote{Note that we do not compare the measured power spectrum to theoretically expected power spectra since the true power in those mocks is not known, except that at large scales where it follows the Kaiser formula given in Eq.~\eqref{eq:Kaiser} with the best-fit parameters from table 1 in \citet{Farr20} for the parameters $\{\ap, \at, b_{\eta,F}, \beta_F, b_{\td,F}\}$, shown in Fig.~\ref{fig:DM_pk}.} 

First, we consider the effects of removing the mean and the slope of each line of sight to mimic the effect of the distortion matrix (dashed-dot lines in Fig.~\ref{fig:mocks_P3D}). This isolates the effect of the distortion matrix modeling and can be seen to reduce the monopole power on all scales with most pronounced impact at large scales where it reduces the power up to a factor of two. The quadrupole is also affected at all scales, leading to an increase in power by approximately 20\%. The hexadecapole is also affected at all scales, and exhibits a sign change at large scales. Note that the effect becomes increasingly small above $k\geq 0.3 \hMpcinv$ for all spectra. 

Second, we assess the impact of the continuum and instrumental noise (dashed lines in Fig.~\ref{fig:mocks_P3D}), denoted by `eboss-0.0', discussed in Sec.~\ref{sec:DM}. This affects all multipoles at large scales. For the monopole it removes, as expected, power at $k\simlt 0.10 \hMpcinv$,  since the large-scale modes are projected out of the continuum. For the quadrupole (red dashed line) it enhances power at all scales with a further enhancement of $10-15\%$ up to $k\simlt 0.15 \hMpcinv$. The hexadecapole is strongly affected and switches sign below $k\simlt 0.20 \hMpcinv$. Note that we do not expect the `raw-dist' mocks to agree exactly with the `0.0' mocks since removing the mean from each line of sight is not exactly the same as introducing a continuum in each spectrum (the latter also affecting the weights). These differences will propagate into the distortions and yield the observed qualitative agreement with a small offset. 

Thirdly (dotted lines in Fig.~\ref{fig:mocks_P3D}), we include metals, HCDs and random redshift errors. For the monopole (dashed to dotted) this adds $\sim10\%$ power at all scales, creating a small off-set. For the quadrupole this mostly affects scales up to $k\simlt 0.15 \hMpcinv$. The hexadecapole, however, is barely affected. We see that the measurements are noisier and that the power is further decreased in two regions around $k\sim 0.05 \hMpcinv$ and $k\sim 0.15 \hMpcinv$. 

\subsection{Anisotropies from the survey geometry}\label{sec:xi_fiducial_test}
\begin{figure}
    \centering
    \includegraphics[width=1\linewidth]{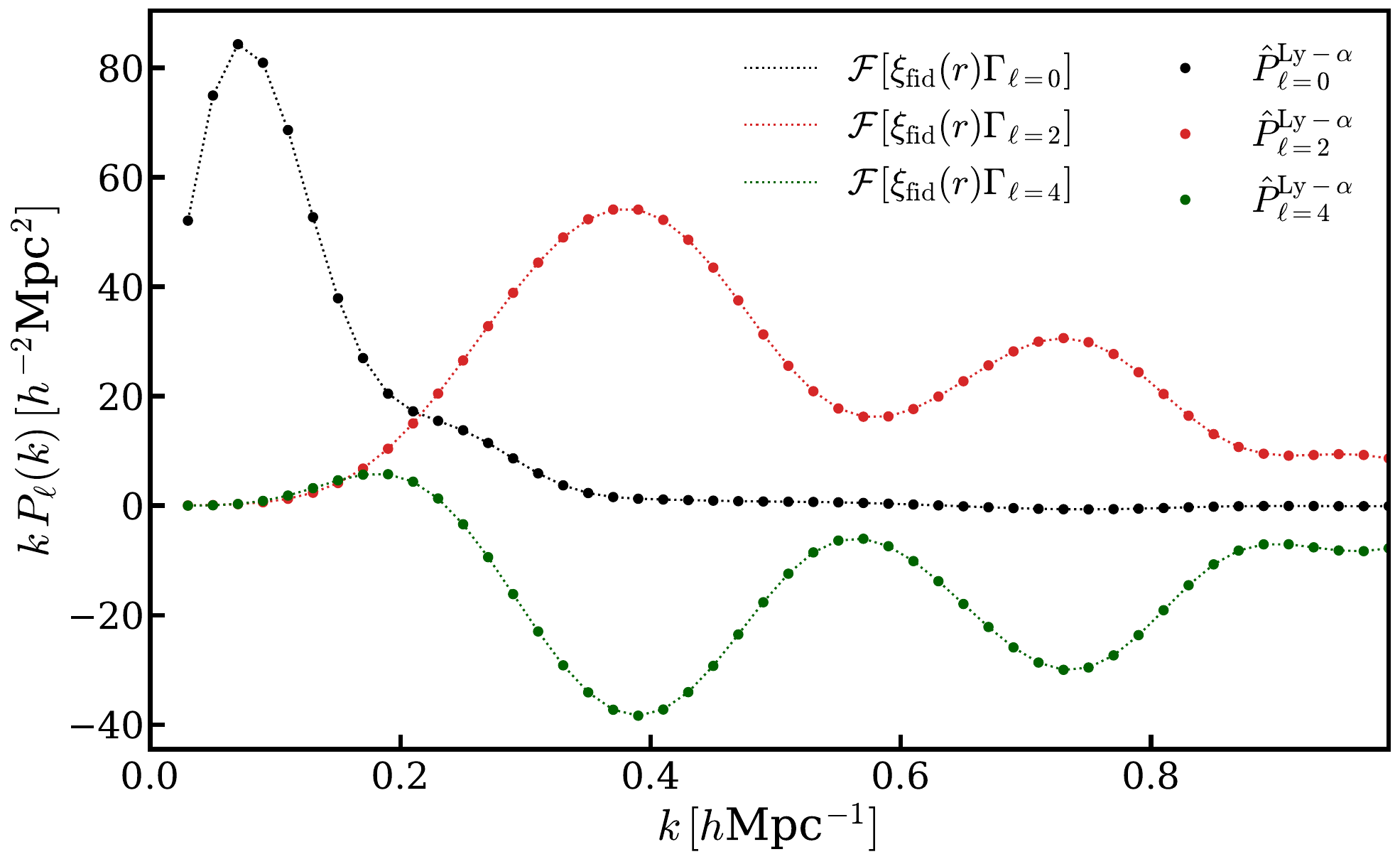}
    \vspace{-0.1in}
    \caption{Consistency test for anisotropies induced by the (eBOSS-like) survey geometry. This shows the power spectra computed from an isotropic fiducial correlation function (without stochasticity), estimated from skewers taken from a \Lyacolore mock using pairs up to $R_0=200\hinvMpc$. The dots represent the measurement of the P3D estimator (black, red and green for $\ell=0,2,4$, respectively). The dashed lines are the fiducial correlation function, $\xi(r)$ multiplied by the window function and the pair count separation window, $\Phi_{\ell}W(r;R_0)$ (same color coding). For ease of comparison, we divide the monopole by a factor of 10.}
    \label{fig:xi_fid_eboss}
    \vspace{-0.1in}
\end{figure}

Finally, we illustrate the effects of window function-induced anisotropy by constructing a simple test built upon a known isotropic correlation function, $\xi_{\rm fid}(\vr) = \exp{(-|\vr|/10)}$. To remove the stochasticity of the data set, we replace the $w_iw_j\delta_i\delta_j$ term in our pair count estimator (given in Eq.~\eqref{eq:DD_pk}) with the explicit correlation function $\xi_{\rm fid}$, evaluated at the pair separation $\vr_{ij}=\vr_i-\vr_j$. We apply the resulting estimator to the \Lyacolore mocks as before, which tests the sensitivity of the present approach to recovering anisotropies purely sourced by the survey geometry, \ie, our measurement of the window matrix. Although the input correlation function is isotropic (with $\xi_{\ell>0}=0$), mode-mixing is expected to occur due to the window; thus $\hat{P}_\ell(k)$ will be non-zero in practice. To compare data and theory, we use Eqs.~\eqref{eq:conv1}-\eqref{eq:conv3}, as before, setting $\xi_{\ell>0}=0$. In this limit, $\xi_{\ell}(r)=\xi_0(r)\Gamma^2{\ell}(r)$.\footnote{In the limit of a fully sampled and infinite box the recovered signal would be isotropic with $\Hat{P}_{\ell=2}=\Hat{P}_{\ell=4}=0$.}

In Fig.~\ref{fig:xi_fid_eboss} we show the results using all the skewers from our `eboss-0.2' mocks which mimic the realistic eBOSS survey geometry.\footnote{We performed the same test using the eBOSS DR16 skewers and recover the anisotropies to 0.1\% precision over the same $k$-range.} The $\td$-fields of these mocks have HCDs that are masked out, \ie this introduces an additional level of realism in reconstructing the window function. We recover the monopole (black), quadrupole (red) and hexadecapole (green) to better than 0.1\% precision over the entire $k$-range of $(0.0 < k \leq 1.0) \hMpcinv$, indicating that our modeling of the window function is highly accurate. This gives us confidence in the present approach to measure the three-dimensional power spectrum at all scales from \Lya data. 

\section{Results} \label{sec:results}
\begin{figure*}
    \begin{minipage}{0.5\linewidth}
        \centering
        \includegraphics[width=1\linewidth]{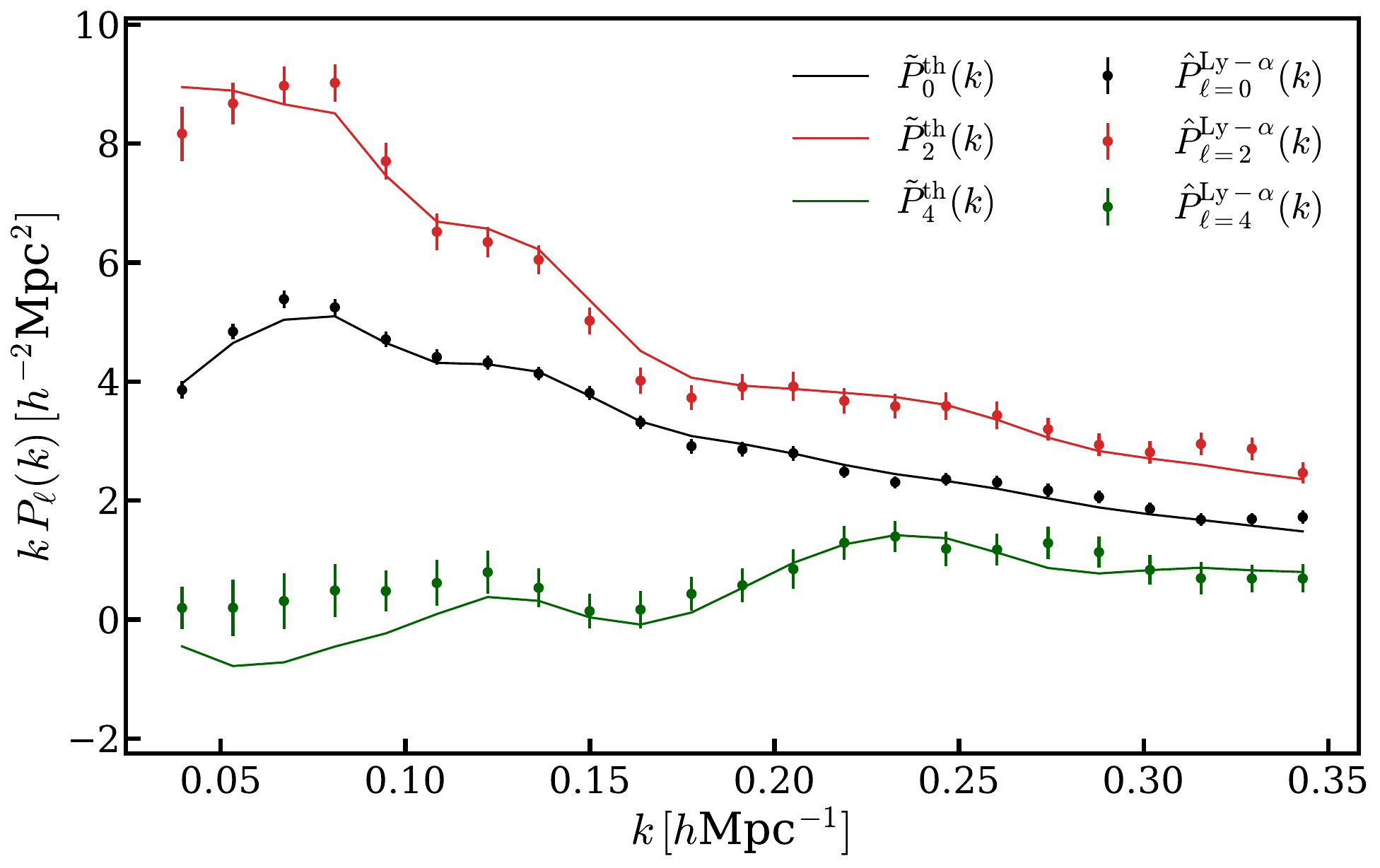}
    \end{minipage}%
    \begin{minipage}{0.5\linewidth}
        \centering
        \includegraphics[width=1\linewidth]{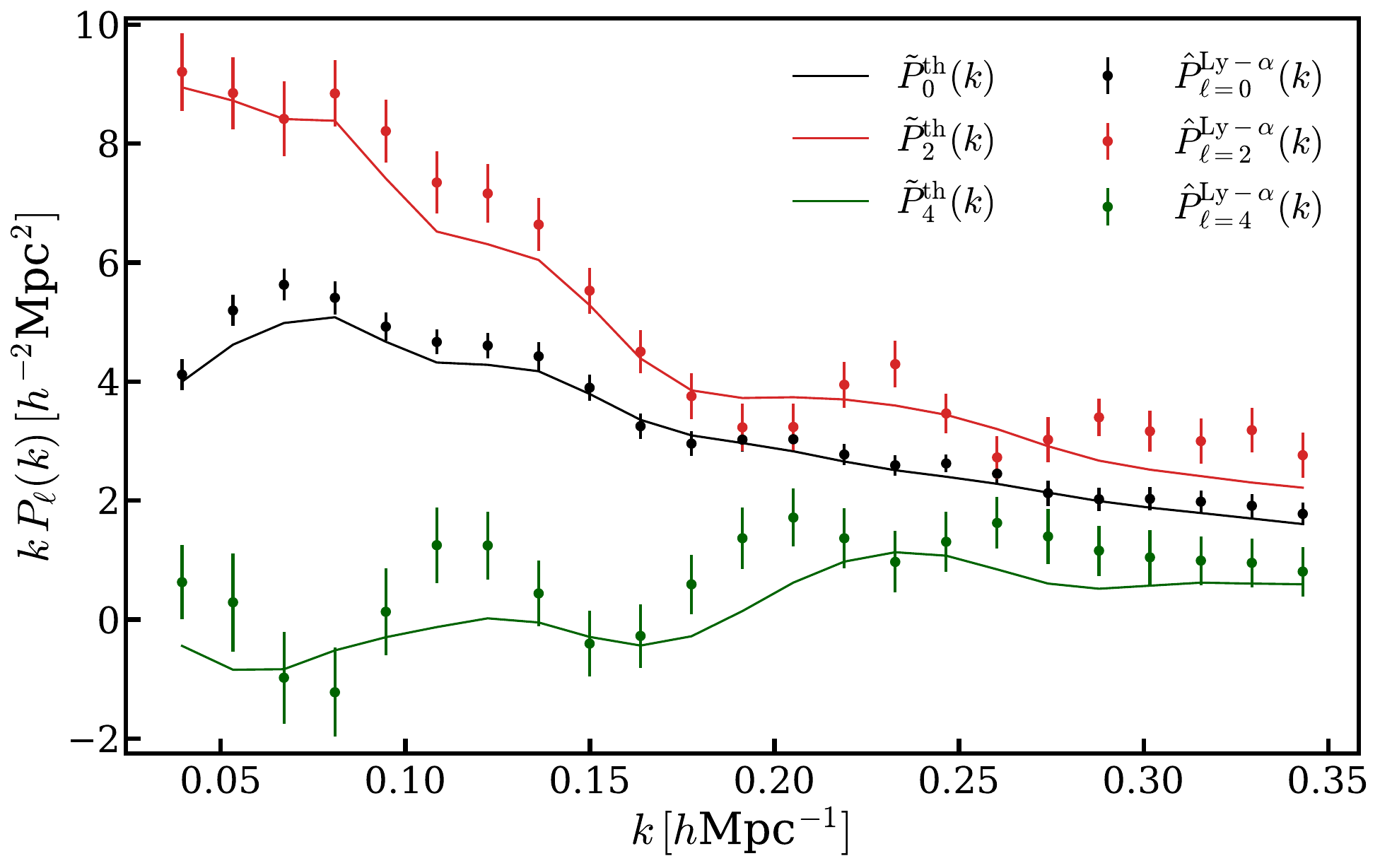}
    \end{minipage}
    \caption{Power spectrum multipoles measured from eBOSS DR16 data for the NGC (left) and SGC (right), showing the monopole in black, quadrupole in red, hexadecapole in green. The dots give the measurements and the solid lines show the quasi-linear theory predictions using eBOSS DR16 best-fit values \citepalias{dMdB:2020}. The error bars are the square root of the diagonal of the covariance matrix obtained from $N=100$ realistic \Lyacolore mocks. The window function contributions are computed up to $\ell_{\rm max}=8$ and convolved with the distorted theory power spectrum. All measurements use the pair-count estimator of Eq.~\eqref{eq:DD_pk}, truncated at $R_0=200\hinvMpc$ via a continuous window function $W(r;R_0)$.}
    \label{fig:P3D_eboss}
    \vspace{-0.1in}
\end{figure*}

In this section we use the pair-count estimator described in Sec.~\ref{sec:methods} to measure the three-dimensional power spectrum from 205,012 eBOSS DR16  quasar spectra. In Sec.~\ref{sec:p3d_eboss} and Fig.\,\ref{fig:P3D_eboss} we present the key result of the paper: the P3D measurement on eBOSS data in a single redshift bin $2\leq z \leq 4$ for both data sets (NGC and SGC). We compare the measurements to window convolved theory power spectra described in Sec.~\ref{sec:theory_pk}. Throughout this section we adopt a \Planck 2016 best-fit $\Lambda$CDM cosmology: $\Omega_{\rm c}h^2  = 0.1197\ ,  \Omega_{\rm b}h^2 = 0.02222\ ,  \Omega_{\nu}h^2 = 0.0006\ ,H_0  = 67.31 \Hunit \ , \sigma_8 = 0.8299\ ,  n_{\rm s} = 0.9655$ \citep{2016A&A...594A..13P}. For the \Lya forest, we use an effective mean redshift of $z=2.334$ and for theory power spectra, given in Eq.~\eqref{eq:Kaiser}, we use the eBOSS DR16 best-fit values from the \Lya auto-correlation presented in table 6 in \citetalias{dMdB:2020}. The BAO parameters are defined as usual \beqa \ap &\equiv \frac{D_H(z_{\rm eff})/r_d}{\left[D_H(z_{\rm eff})/r_d\right]_{\rm fid}} = 1.047\ , \\  \at &\equiv \frac{D_M(z_{\rm eff})/r_d}{\left[D_M(z_{\rm eff})/r_d\right]_{\rm fid}} = 0.980\ , \eeqa where $D_H$ and $D_M$ are the Hubble and comoving angular diameter distances, respectively, evaluated at some effective redshift, $z_{\rm eff}$. The denominator is evaluated at a fiducial $\Lambda$CDM cosmology. The bias and RSD best-fit parameters are \citepalias{dMdB:2020} \beqa b_{\eta,\, \text{\Lya}} & = -0.201, & \beta_{\text{\Lya}} & = 1.657\ , \eeqa the metal biases with the associated restframe wavelength in parentheses are \beqa b_{\eta,\, \text{\siii(1190)}} & = -0.0029\ , & b_{\eta,\, \text{\siii(1193)}} & = -0.0021\ , \\ b_{\eta,\, \text{\siiii(1207)}} & = -0.0045\ , & b_{\eta,\, \text{\siii(1260)}} & = -0.0022\ , \eeqa  and the bias parameters for the HCD systems \citep{Rogers:2018} \beqa b_{\rm HCD} & = -0.0522\ , & \beta_{\rm HCD} & = 0.610\ .\eeqa
In Sec.~\ref{sec:cov_mat} we discuss our error bars and covariance matrix for the P3D measurement obtained from $N=100$ realistic \Lya simulations. Increasing the pair separation in the pair truncation window function $W(r;R_0)$ to $R_0= 400 \hinvMpc$ did not improve the agreement between theory and measured power spectrum, thus we do not show the corresponding results in this work. For our main analysis, we measure the power spectra up to $\ell_{\rm max}=4$ in $24$ equidistant $k$-bins in the range $(0.02 \leq k \leq 0.35) \hMpcinv$ and the window function multipoles up to $\ell_{\rm max}=8$ with $N_r=1,000$ linearly spaced bins in the range $(0 \leq r \leq R_0=200) \hMpcinv$, respectively. 

\subsection{Power spectrum multipoles from eBOSS DR16}\label{sec:p3d_eboss}
In Fig.~\ref{fig:P3D_eboss} we show the three dimensional power spectrum multipoles measured on eBOSS DR16 data for the NGC (SGC) in the left (right) panel and compare them to the window convolved quasi non-linear theory power spectrum multipoles. We show the monopole, quadrupole and hexadecapole from the data and compute the window function multipoles up to $\ell_{\rm max}=8$, as described in Sec.~\ref{sec:theory_pk}. The error bars are given by the square root of the diagonal of the covariance matrix, introduced in  Sec.~\ref{sec:cov_mat}. To account for the employed continuum fitting procedure that projects out large-scale modes, we apply a distortion matrix, further altering the large scale shape.\footnote{We remind the reader that we use the same data selection and \Lya flux decrement obtained from eBOSS DR16 data for our analysis as~\citetalias{dMdB:2020}.} The estimator uses a pair separation window truncating the pair count at $R_0\geq200 \hinvMpc$, as before. 

For the monopole, we find excellent agreement with the theory prediction over the entire $k$-range. The quadrupole is also in good agreement with the theory prediction, bar the last few measurement points in the $k\simgt 0.20 \hMpcinv$ range. The quadrupole and hexadecapole both have correlated features at large scales: We conjecture that this stems from the mode mixing introduced by the distortion matrix. In particular, the $\sim 1-2\sigma$ discrepancy at large scales in the hexadecapole has to be dealt with cautiously when extracting cosmology from the \Lya forest. In Sec.~\ref{sec:DM} and in Fig.~\ref{fig:DM_pk}, we justify this by showing that the forward modeling of the DM does not capture exactly the hexadecapole on large scales: First, the distortion matrix is computed out to separations of $\simgt 400 \hinvMpc$ but should, in principle, be computed up to the scales of the survey. Second, it assumes a correlation function measurement in bins of size $\{4\hinvMpc \times \rpar,4\hinvMpc \times \rperp\}$. In the range  $k\simgt 0.2 \hMpcinv$, the quadrupole and hexadecapole measurements and the theory predictions are in $\sim 1\sigma$ agreement with each other. Note that the error bars from SGC are larger than the ones from the NGC which stems from the smaller statistics, \ie, NGC approximately has three times the number of spectra than the SGC data set.

\begin{figure*}
    \begin{minipage}{0.5\linewidth}
        \centering
        \includegraphics[width=1\linewidth]{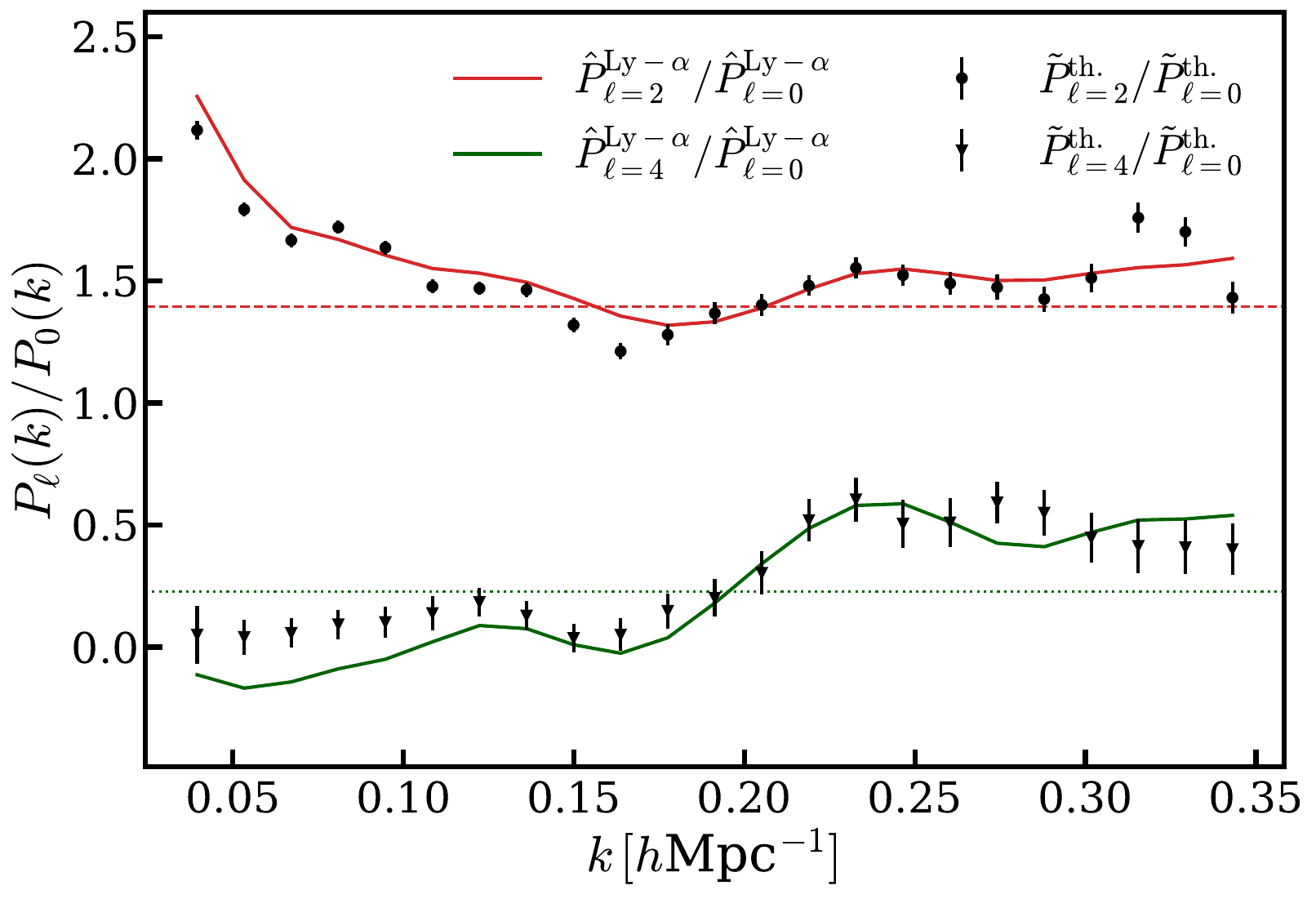}
    \end{minipage}%
    \begin{minipage}{0.5\linewidth}
        \centering
        \includegraphics[width=1\linewidth]{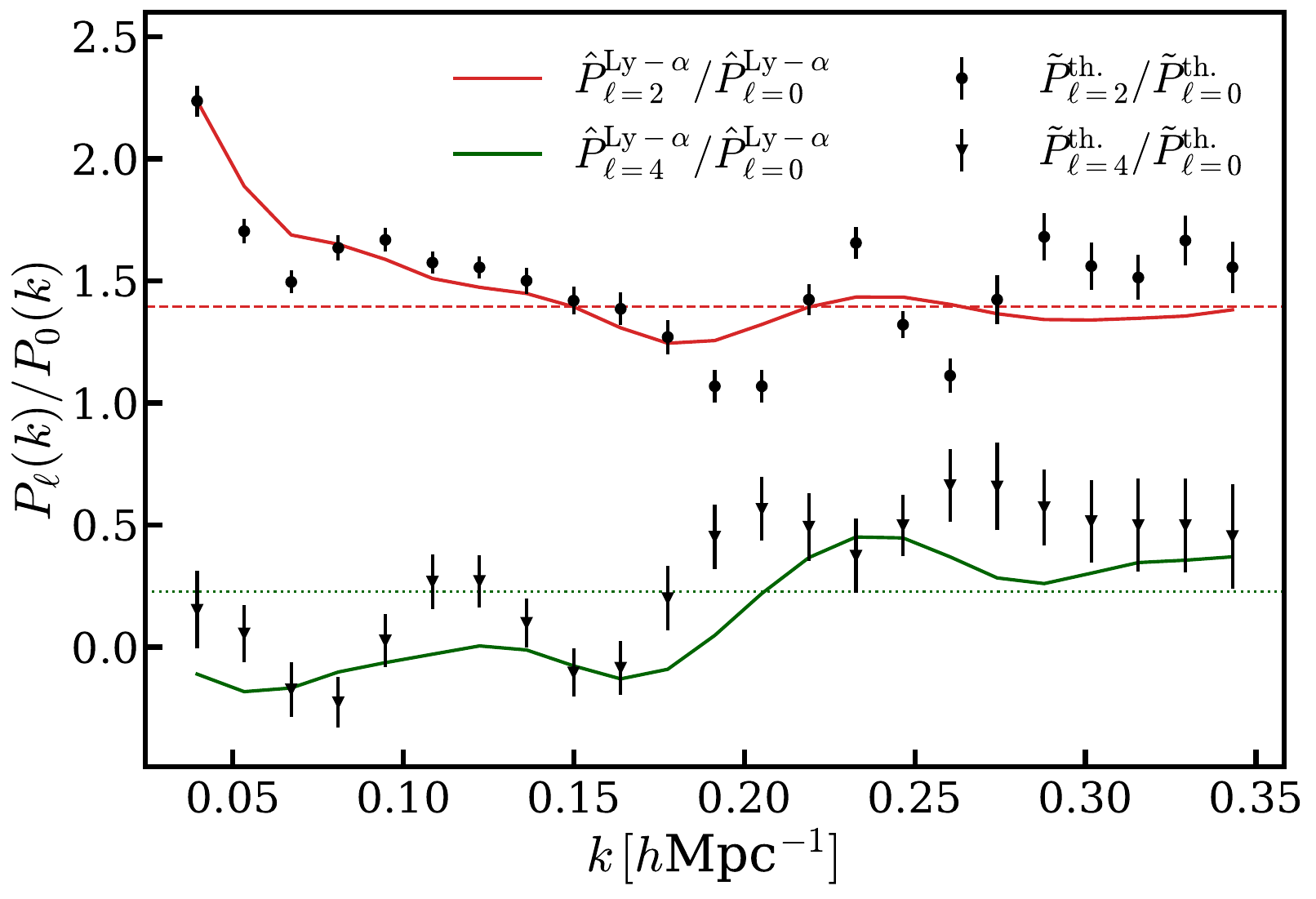}
    \end{minipage}
    \caption{Ratio of measured quadrupole (black dots) and hexadecapole (black inverted triangles) power spectra to monopole power for eBOSS DR16 data (left panel: NGC, right panel: SGC ), with linear expectation for $\beta_F = 1.627$ given by the horizontal dashed lines and ratio of the (non-linear) theory power spectra for the ratio of the quadrupole (red line) and hexadecapole (green line) to the theory monopole. The error bars are taken from the power spectrum, \ie,~the square root of the diagonal of the covariance matrix obtained from \Lyacolore simulations, and are rescaled by the monopole power.}
    \label{fig:ratio_Pk_eboss}
    \vspace{-0.1in}
\end{figure*}

In Fig.~\ref{fig:ratio_Pk_eboss} we show the ratio of the quadrupole and hexadecapole to the monopole to obtain an estimate for the RSD parameter, $\beta_F$. Note that although the measurements are quite noisy, we get good agreement for the measured power spectra to the non-linear theory predictions at scales $k\simgt 0.1 \hMpcinv$. For comparison, we include the linear theory prediction (horizontal dashed lines) to guide the eye which we compute by integrating Eq.~\eqref{eq:RSD} in Fourier space. Assuming non-linear effects are negligible, the Kaiser effect gives the analytic prefactors $Q_{\ell=0}(\beta)=1+2\beta/3+\beta^2/5$, $Q_{\ell=2}(\beta)=4\beta/3+4\beta^2/7$, and $Q_{\ell=4}(\beta)=8\beta^2/35$ for the monopole, quadrupole and hexadecapole, respectively. In this limit, the multipoles of our theory power spectrum are given by $P_{\ell}(k) = Q_{\ell}(\beta)b^2P_{\rm lin}(k)$; inserting the value of $\beta = 1.627$ gives 1.39 and 0.23 for the ratios of quadrupole to monopole and hexadecapole to monopole, respectively. The errorbars are taken from the diagonals of the covariance matrices, rescaled by the monopole power, and should be taken indicatively rather than at face value. 

\subsection{Covariance matrix from mocks catalogs} \label{sec:cov_mat}
\begin{figure*}
    \begin{minipage}{0.5\linewidth}
        \centering
            \includegraphics[width=1\linewidth]{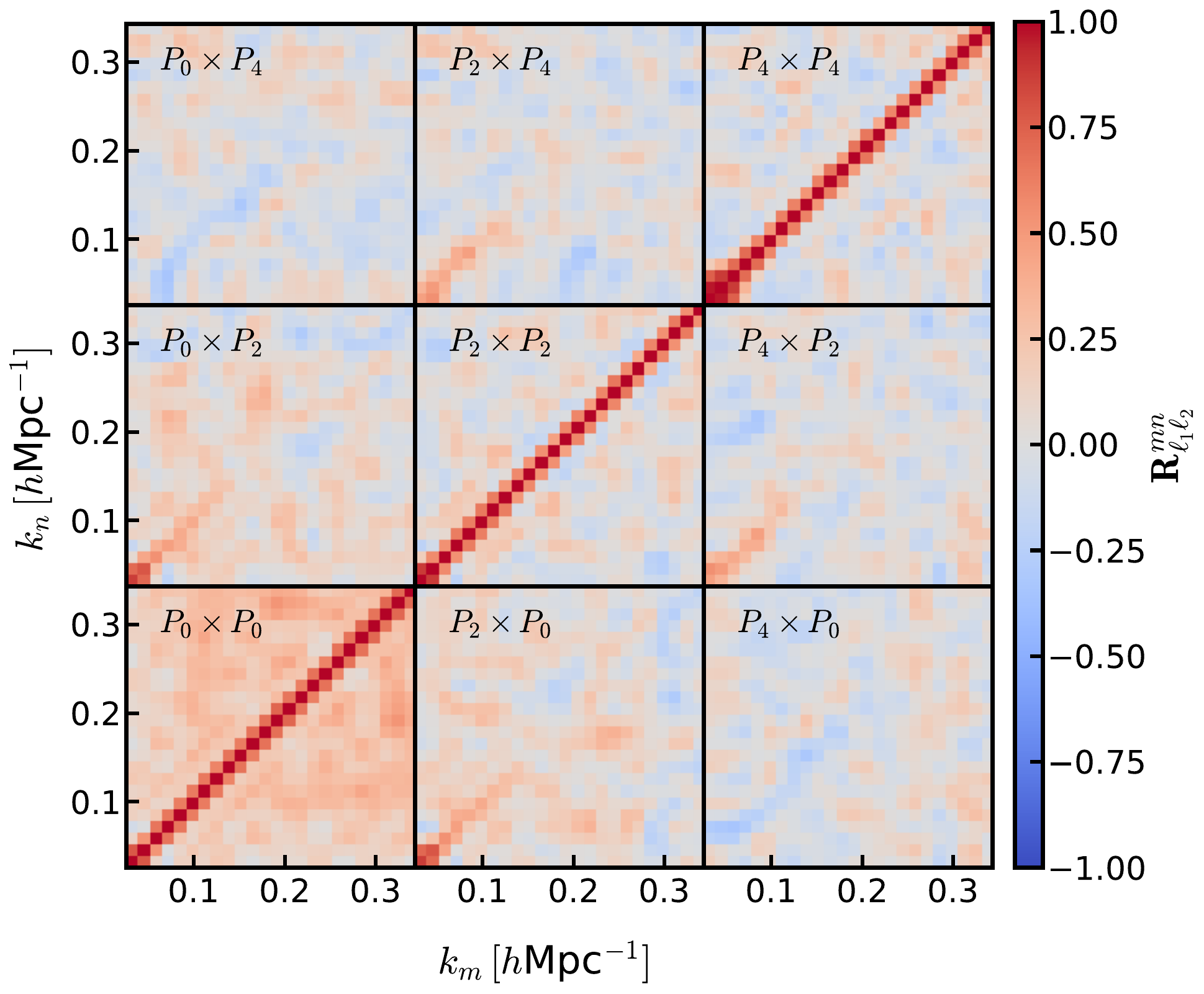}
    \end{minipage}%
    \begin{minipage}{0.5\linewidth}
        \centering
            \includegraphics[width=1\linewidth]{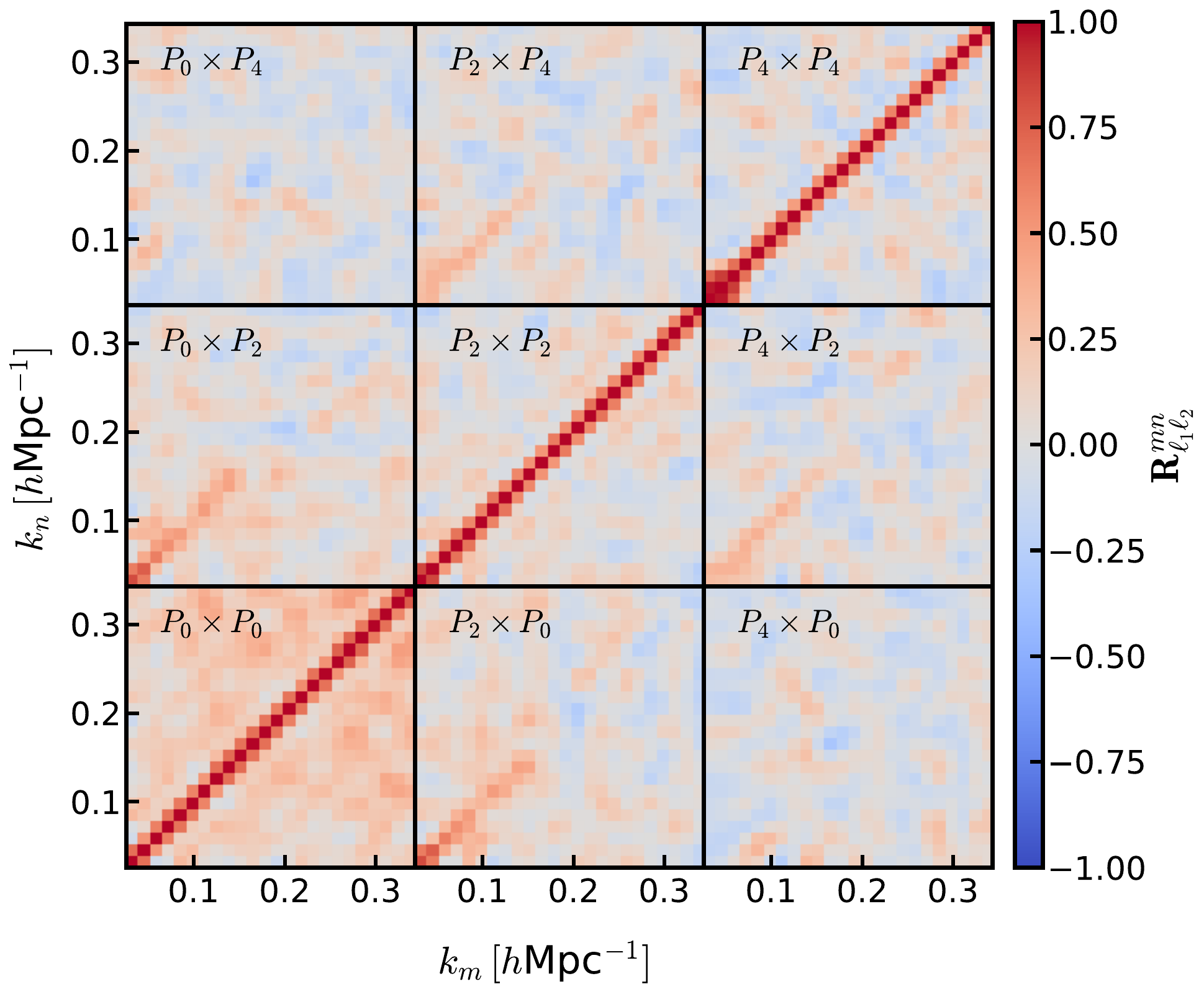}
    \end{minipage}
    \caption{Correlation matrix of the power spectrum multipoles $P_{\ell_1}\times P_{\ell_2}$ obtained from $N=100$ realizations of the \Lyacolore simulations, emulating that of the eBOSS data. The left (right) panel shows the NGC (SGC) measurements. For each multipole, we display the $k$-modes used in the analysis: $0.02 \leq k \leq 0.35 \hMpcinv$. Pair counts are truncated above $R_0 > 200\hinvMpc$. The correlation matrix is defined in Eq.~\eqref{eq:corr_mat}; red denotes fully correlated and blue fully anti-correlated power spectrum bins. In general, the correlation matrix shows little off-diagonal terms except for the anisotropies sourced by the sparse \Lya sampling which are visible in the `cross' terms $P_{\ell}\times P_{\ell'}$ for the combinations $\{\ell=0,\ell'=2\}$ and $\{\ell=2,\ell'=4\}$ (and their inverses). The auto-correlation of the monopole is at the level of 10-30\% for the off-diagonal terms. Cross-terms of the hexadecapole with the monopole are uncorrelated and with the quadrupole only slightly correlated.}
    \label{fig:corr_mat_eboss_lyacolore}
    \vspace{-0.1in}
\end{figure*}

To capture the variance between spectra and account for instrumental and systematic noise in the data, we compute a covariance matrix from $N=100$ realizations of `eboss-0.2' \Lyacolore simulations, introduced in Sec.~\ref{sec:lyacolore}. The error bars in Fig.~\ref{fig:P3D_eboss} are the square root of the diagonal of the covariance matrix with the corresponding correlation matrix shown in Fig.~\ref{fig:corr_mat_eboss_lyacolore} for which we vary both Legendre multipoles and $k$-bins, as in Sec.~\ref{sec:GRF}. The resulting correlation matrix has a diagonal structure with non-negligible off-diagonal terms and blocks: It is interesting to note that the eBOSS DR16 survey geometry mixes modes between `neighboring multipoles', e.g., between monopole and quadrupole as well as between quadrupole and hexadecapole. For the $\ell_1=\ell_2$ correlations, we observe a correlation length of up to two $k$-bins. Note that we do not include any scales with $k^{-1}$ larger than the pair separation window $R_0$, thus (in contrast to the GRF correlation matrix) the lowest $k$-bins are not artificially correlated. The resulting covariance matrix is positive definite and can thus be used for cosmological inference. We note that the covariance matrix of the monopole, $P_{\ell=0}\times P_{\ell=0}$, shows correlations between $k$-bins of order 10\% on larger and up to 30\% on smaller scales. This may stem from the application of the distortion matrix, which effectively mixes $\td$-modes. Thus, we expect a less correlated correlation matrix when using an alternative continuum fitting method that does not project out large-scale modes at the expense of increased noise in the forest.

\section{Summary and Conclusions} \label{sec:conclusions}
The \Lya forest is a treasure trove of cosmological information on the expansion history of the Universe and beyond. Given its high-redshift range ($2\leq z \leq 4$) and sensitivity to Mpc scales and below, it is an ideal tracer to probe a wide range of scales: early-Universe physics from combinations with a large-scale tracer, \eg, the cosmic microwave background anisotropies, from large scales or dark matter models as well as thermal properties of the ionized (cold) IGM from small scales. Previous analyses have focused on the two-point correlation function \citep[see, e.g.,~][]{dMdB:2020} or the small-scale one-dimensional power spectrum from Fourier transforms along each line of sight \citep[see, e.g.,~][in the context of DESI]{2023MNRAS.526.5118R,2024MNRAS.tmp..176K}; however, a pioneering study on simulations \citep{Font-Ribera:2018} and a recent small-scale measurement \citep{karim2023measurement} paired with advancements in the theoretical modeling of the \Lya forest \citep[see, e.g.,~][]{Ivanov:2023} have now paved the way for three-dimensional power spectrum analyses. To achieve this goal, we require accurate estimates of the 3D \Lya power spectrum: this is challenging, given the sparse sampling transverse to and dense sampling along the line-of-sight. In this paper, we attack this problem by presenting a pair count estimator to measure the three-dimensional power spectrum from the \Lya forest. Our approach is based on the pair counting estimator \texttt{HIPSTER} \citep{Philcox:2021, Philcox2020} and weighs each pair by $\exp(i\mat{k}\cdot \mat{r}_{ij})$, for wave vector $\mat k$ and particle pair separation $\mat{r}_{ij}=\mat{r}_i-\mat{r}_j$. This directly measures the power spectrum without the need for grid-based Fourier transforms which are affected by the sparse $\delta-$function type sampling of the \Lya forest. 

We have presented the first large-scale three-dimensional power spectrum measurement of Lyman-$\alpha$ forest spectra on 205,012 medium-resolution \Lya forest spectra from the extended Baryon Oscillation Spectroscopic Survey (eBOSS) DR16 in two disjoint regions of the sky: the northern and southern galactic cap. Furthermore, we have compared our P3D measurement to the best-fit quasi-linear theory prediction from the 2PCF presented in \citetalias{dMdB:2020}. We extensively test the estimator on Gaussian random fields, \ie, realizations of a linear input power spectrum, with Gaussian noise and realistic continuum error levels of $10\%$. We recover the input power spectrum at the $\sim (1)~3\sigma$ level for (dense) sparse configurations, respectively. Further, our pipeline has been extensively tested on synthetic \Lya spectra (\Lyacolore) with increasing levels of realism to probe for the effects of distorted continua, instrumental noise, HCDs, metals and random redshift errors. We demonstrated that we can forward model the effects of anisotropies sourced by the survey geometry and the distortions introduced by the continuum fitting (modes along the line-of-sight, $\kpar=0$) accurately for the multipoles. 

We present a covariance matrix derived from 100 realistic \Lyacolore simulations for further cosmological analysis of the \Lya forest data set. We obtain a fairly diagonal covariance matrix with correlations between adjacent multipoles stemming from the survey geometry and, for the auto-correlation of the monopole, contaminants and the mode-mixing distortion matrix. We present the anisotropic clustering of eBOSS DR16 \Lya spectra up to the hexadecapole and obtain good agreement to the Kaiser formula with a non-linear correction term obtained from hydrodynamical simulations \citep{Arinyo:2015} up to $k \leq 0.35 \hMpcinv$. The best-fit theory power spectrum obtained from \citetalias{dMdB:2020} shows deviations at the $\sim 2\sigma$ level on the largest scales, in particular for the hexadecapole. The quadrupole and  hexadecapole mix a large range of scales when convolving the window matrix with the correlation function and Hankel transforming it to Fourier space. This effect is even more pronounced if a distortion matrix (an artifact from the present continuum fitting approach) is applied. We leave for future work to explore alternative continuum fitting methods as well as the accuracy of the presented covariance matrix. 

Our main conclusion is that the novel estimator is well suited to measure clustering statistics from the \Lya forest and can deal with non-trivial survey geometries and masked data vectors. This will facilitate robust measurements of P3D for the currently observing Dark Energy Spectroscopic Instrument and future surveys such as the Prime Focus Spectrograph, WEAVE-QSO and 4MOST, opening the door to a wide variety of novel cosmological analyses.

\section*{Acknowledgments}
The authors are indebted to M.~White, A.~Font-Ribera, and A.~Cuceu for fruitful discussions and to A.~Slosar for motivating this work. We are grateful to the members of the SDSS/eBOSS collaboration for their enormous efforts in producing such a wonderful data set. The authors thank A.~X.~Gonz\'{a}lez-Morales, H.~K.~Herrera-Alcantar, C.~Ram\'irez-P\'erez, and A.~Mu\~{n}oz-Guti\'errez for their work on the eBOSS \Lyacolore mocks and B.~Abolfathi for extending \texttt{specsim} to support eBOSS simulations. 

This research used resources supported by the U.S. Department of Energy (DOE), Office of Science, Office of High-Energy Physics, under Contract No. DE-AC02-05CH11231, and by the National Energy Research Scientific Computing Center, a DOE Office of Science User Facility under the same contract. OHEP is a Junior Fellow of the Simons Society of Fellows, and completed much of this work on the beach in Mexico. VI acknowledges support by the Kavli Foundation.

Funding for the Sloan Digital Sky 
Survey IV has been provided by the 
Alfred P. Sloan Foundation, the U.S. 
Department of Energy Office of 
Science, and the Participating 
Institutions. SDSS-IV acknowledges support and 
resources from the Center for High 
Performance Computing  at the 
University of Utah. The SDSS 
website is \url{www.sdss4.org}. SDSS-IV is managed by the 
Astrophysical Research Consortium 
for the Participating Institutions 
of the SDSS Collaboration including 
the Brazilian Participation Group, 
the Carnegie Institution for Science, 
Carnegie Mellon University, Center for 
Astrophysics | Harvard \& 
Smithsonian, the Chilean Participation 
Group, the French Participation Group, 
Instituto de Astrof\'isica de 
Canarias, The Johns Hopkins 
University, Kavli Institute for the 
Physics and Mathematics of the 
Universe (IPMU) / University of 
Tokyo, the Korean Participation Group, 
Lawrence Berkeley National Laboratory, 
Leibniz Institut f\"ur Astrophysik 
Potsdam (AIP),  Max-Planck-Institut 
f\"ur Astronomie (MPIA Heidelberg), 
Max-Planck-Institut f\"ur 
Astrophysik (MPA Garching), 
Max-Planck-Institut f\"ur 
Extraterrestrische Physik (MPE), 
National Astronomical Observatories of 
China, New Mexico State University, 
New York University, University of 
Notre Dame, Observat\'ario 
Nacional / MCTI, The Ohio State 
University, Pennsylvania State 
University, Shanghai 
Astronomical Observatory, United 
Kingdom Participation Group, 
Universidad Nacional Aut\'onoma 
de M\'exico, University of Arizona, 
University of Colorado Boulder, 
University of Oxford, University of 
Portsmouth, University of Utah, 
University of Virginia, University 
of Washington, University of 
Wisconsin, Vanderbilt University, 
and Yale University.

\section*{Data Availability}
The data underlying this article will be shared on reasonable request. The eBOSS DR16 data set is publicly available at \url{https://data.sdss.org/sas/dr16}. 


\bibliographystyle{mnras}
\bibliography{references} 

\begin{thebibliography}{}
\makeatletter
\relax
\def\mn@urlcharsother{\let\do\@makeother \do\$\do\&\do\#\do\^\do\_\do\%\do\~}
\def\mn@doi{\begingroup\mn@urlcharsother \@ifnextchar [ {\mn@doi@}
  {\mn@doi@[]}}
\def\mn@doi@[#1]#2{\def\@tempa{#1}\ifx\@tempa\@empty \href
  {http://dx.doi.org/#2} {doi:#2}\else \href {http://dx.doi.org/#2} {#1}\fi
  \endgroup}
\def\mn@eprint#1#2{\mn@eprint@#1:#2::\@nil}
\def\mn@eprint@arXiv#1{\href {http://arxiv.org/abs/#1} {{\tt arXiv:#1}}}
\def\mn@eprint@dblp#1{\href {http://dblp.uni-trier.de/rec/bibtex/#1.xml}
  {dblp:#1}}
\def\mn@eprint@#1:#2:#3:#4\@nil{\def\@tempa {#1}\def\@tempb {#2}\def\@tempc
  {#3}\ifx \@tempc \@empty \let \@tempc \@tempb \let \@tempb \@tempa \fi \ifx
  \@tempb \@empty \def\@tempb {arXiv}\fi \@ifundefined
  {mn@eprint@\@tempb}{\@tempb:\@tempc}{\expandafter \expandafter \csname
  mn@eprint@\@tempb\endcsname \expandafter{\@tempc}}}

\bibitem[\protect\citeauthoryear{{Abareshi} et~al.,}{{Abareshi}
  et~al.}{2022}]{DESI:2022}
{Abareshi} B.,  et~al., 2022, arXiv e-prints, \href
  {https://ui.adsabs.harvard.edu/abs/2022arXiv220510939A} {p. arXiv:2205.10939}

\bibitem[\protect\citeauthoryear{{Adame} et~al.,}{{Adame}
  et~al.}{2024}]{2024AJ....167...62A}
{Adame} A.~G.,  et~al., 2024, \mn@doi [\aj] {10.3847/1538-3881/ad0b08}, \href
  {https://ui.adsabs.harvard.edu/abs/2024AJ....167...62A} {167, 62}

\bibitem[\protect\citeauthoryear{{Afshordi}, {McDonald}  \&
  {Spergel}}{{Afshordi} et~al.}{2003}]{Afshordi:2003}
{Afshordi} N.,  {McDonald} P.,   {Spergel} D.~N.,  2003, \mn@doi [\apjl]
  {10.1086/378763}, \href
  {https://ui.adsabs.harvard.edu/abs/2003ApJ...594L..71A} {594, L71}

\bibitem[\protect\citeauthoryear{{Alam} et~al.,}{{Alam}
  et~al.}{2021}]{Alam:2021}
{Alam} S.,  et~al., 2021, \mn@doi [\prd] {10.1103/PhysRevD.103.083533}, \href
  {https://ui.adsabs.harvard.edu/abs/2021PhRvD.103h3533A} {103, 083533}

\bibitem[\protect\citeauthoryear{{Arinyo-i-Prats}, {Miralda-Escud{\'e}}, {Viel}
   \& {Cen}}{{Arinyo-i-Prats} et~al.}{2015}]{Arinyo:2015}
{Arinyo-i-Prats} A.,  {Miralda-Escud{\'e}} J.,  {Viel} M.,   {Cen} R.,  2015,
  \mn@doi [\jcap] {10.1088/1475-7516/2015/12/017}, \href
  {https://ui.adsabs.harvard.edu/abs/2015JCAP...12..017A} {2015, 017}

\bibitem[\protect\citeauthoryear{{Armengaud}, {Palanque-Delabrouille},
  {Y{\`e}che}, {Marsh}  \& {Baur}}{{Armengaud} et~al.}{2017}]{Armengaud:2017}
{Armengaud} E.,  {Palanque-Delabrouille} N.,  {Y{\`e}che} C.,  {Marsh} D.
  J.~E.,   {Baur} J.,  2017, \mn@doi [\mnras] {10.1093/mnras/stx1870}, \href
  {https://ui.adsabs.harvard.edu/abs/2017MNRAS.471.4606A} {471, 4606}

\bibitem[\protect\citeauthoryear{{Baur}, {Palanque-Delabrouille}, {Y{\`e}che},
  {Magneville}  \& {Viel}}{{Baur} et~al.}{2016}]{Baur:2016}
{Baur} J.,  {Palanque-Delabrouille} N.,  {Y{\`e}che} C.,  {Magneville} C.,
  {Viel} M.,  2016, \mn@doi [\jcap] {10.1088/1475-7516/2016/08/012}, \href
  {https://ui.adsabs.harvard.edu/abs/2016JCAP...08..012B} {2016, 012}

\bibitem[\protect\citeauthoryear{{Beutler} \& {McDonald}}{{Beutler} \&
  {McDonald}}{2021}]{2021JCAP...11..031B}
{Beutler} F.,  {McDonald} P.,  2021, \mn@doi [\jcap]
  {10.1088/1475-7516/2021/11/031}, \href
  {https://ui.adsabs.harvard.edu/abs/2021JCAP...11..031B} {2021, 031}

\bibitem[\protect\citeauthoryear{{Beutler} et~al.,}{{Beutler}
  et~al.}{2017}]{Beutler:2016}
{Beutler} F.,  et~al., 2017, \mn@doi [\mnras] {10.1093/mnras/stw3298}, \href
  {https://ui.adsabs.harvard.edu/abs/2017MNRAS.466.2242B} {466, 2242}

\bibitem[\protect\citeauthoryear{Blomqvist et~al.,}{Blomqvist
  et~al.}{2018}]{Blomqvist_2018}
Blomqvist M.,  et~al., 2018, \mn@doi [Journal of Cosmology and Astroparticle
  Physics] {10.1088/1475-7516/2018/05/029}, 2018, 029

\bibitem[\protect\citeauthoryear{{Boera}, {Becker}, {Bolton}  \&
  {Nasir}}{{Boera} et~al.}{2019}]{Boera:2019}
{Boera} E.,  {Becker} G.~D.,  {Bolton} J.~S.,   {Nasir} F.,  2019, \mn@doi
  [\apj] {10.3847/1538-4357/aafee4}, \href
  {https://ui.adsabs.harvard.edu/abs/2019ApJ...872..101B} {872, 101}

\bibitem[\protect\citeauthoryear{{Bolton}, {Viel}, {Kim}, {Haehnelt}  \&
  {Carswell}}{{Bolton} et~al.}{2008}]{Bolton:2008}
{Bolton} J.~S.,  {Viel} M.,  {Kim} T.~S.,  {Haehnelt} M.~G.,   {Carswell}
  R.~F.,  2008, \mn@doi [\mnras] {10.1111/j.1365-2966.2008.13114.x}, \href
  {https://ui.adsabs.harvard.edu/abs/2008MNRAS.386.1131B} {386, 1131}

\bibitem[\protect\citeauthoryear{{Bolton}, {Puchwein}, {Sijacki}, {Haehnelt},
  {Kim}, {Meiksin}, {Regan}  \& {Viel}}{{Bolton} et~al.}{2017}]{Bolton:2017}
{Bolton} J.~S.,  {Puchwein} E.,  {Sijacki} D.,  {Haehnelt} M.~G.,  {Kim} T.-S.,
   {Meiksin} A.,  {Regan} J.~A.,   {Viel} M.,  2017, \mn@doi [\mnras]
  {10.1093/mnras/stw2397}, \href
  {https://ui.adsabs.harvard.edu/abs/2017MNRAS.464..897B} {464, 897}

\bibitem[\protect\citeauthoryear{{Busca} et~al.,}{{Busca}
  et~al.}{2013}]{Busca:2013}
{Busca} N.~G.,  et~al., 2013, \mn@doi [\aap] {10.1051/0004-6361/201220724},
  \href {https://ui.adsabs.harvard.edu/abs/2013A&A...552A..96B} {552, A96}

\bibitem[\protect\citeauthoryear{{Castorina} \& {White}}{{Castorina} \&
  {White}}{2018}]{Castorina2018}
{Castorina} E.,  {White} M.,  2018, \mn@doi [\mnras] {10.1093/mnras/sty410},
  \href {https://ui.adsabs.harvard.edu/abs/2018MNRAS.476.4403C} {476, 4403}

\bibitem[\protect\citeauthoryear{{Chabanier} et~al.,}{{Chabanier}
  et~al.}{2019}]{Chabanier:2019}
{Chabanier} S.,  et~al., 2019, \mn@doi [\jcap] {10.1088/1475-7516/2019/07/017},
  \href {https://ui.adsabs.harvard.edu/abs/2019JCAP...07..017C} {2019, 017}

\bibitem[\protect\citeauthoryear{{Chen}, {Vlah}  \& {White}}{{Chen}
  et~al.}{2021}]{Chen:2021}
{Chen} S.-F.,  {Vlah} Z.,   {White} M.,  2021, \mn@doi [\jcap]
  {10.1088/1475-7516/2021/05/053}, \href
  {https://ui.adsabs.harvard.edu/abs/2021JCAP...05..053C} {2021, 053}

\bibitem[\protect\citeauthoryear{{Cieplak} \& {Slosar}}{{Cieplak} \&
  {Slosar}}{2016}]{Cieplak:2016}
{Cieplak} A.~M.,  {Slosar} A.,  2016, \mn@doi [\jcap]
  {10.1088/1475-7516/2016/03/016}, \href
  {https://ui.adsabs.harvard.edu/abs/2016JCAP...03..016C} {2016, 016}

\bibitem[\protect\citeauthoryear{{Croft}, {Weinberg}, {Katz}  \&
  {Hernquist}}{{Croft} et~al.}{1998}]{1998ApJ...495...44C}
{Croft} R. A.~C.,  {Weinberg} D.~H.,  {Katz} N.,   {Hernquist} L.,  1998,
  \mn@doi [\apj] {10.1086/305289}, \href
  {https://ui.adsabs.harvard.edu/abs/1998ApJ...495...44C} {495, 44}

\bibitem[\protect\citeauthoryear{{Cuceu}, {Font-Ribera}, {Joachimi}  \&
  {Nadathur}}{{Cuceu} et~al.}{2021}]{Cuceu:2021}
{Cuceu} A.,  {Font-Ribera} A.,  {Joachimi} B.,   {Nadathur} S.,  2021, \mn@doi
  [\mnras] {10.1093/mnras/stab1999}, \href
  {https://ui.adsabs.harvard.edu/abs/2021MNRAS.506.5439C} {506, 5439}

\bibitem[\protect\citeauthoryear{{Cuceu}, {Font-Ribera}, {Nadathur}, {Joachimi}
   \& {Martini}}{{Cuceu} et~al.}{2023}]{Cuceu:2023}
{Cuceu} A.,  {Font-Ribera} A.,  {Nadathur} S.,  {Joachimi} B.,   {Martini} P.,
  2023, \mn@doi [\prl] {10.1103/PhysRevLett.130.191003}, \href
  {https://ui.adsabs.harvard.edu/abs/2023PhRvL.130s1003C} {130, 191003}

\bibitem[\protect\citeauthoryear{{DESI Collaboration} et~al.,}{{DESI
  Collaboration} et~al.}{2016}]{DESI:2016}
{DESI Collaboration} et~al., 2016, arXiv e-prints, \href
  {https://ui.adsabs.harvard.edu/abs/2016arXiv161100036D} {p. arXiv:1611.00036}

\bibitem[\protect\citeauthoryear{{Dawson} et~al.,}{{Dawson}
  et~al.}{2016}]{Dawson:2016}
{Dawson} K.~S.,  et~al., 2016, \mn@doi [\aj] {10.3847/0004-6256/151/2/44},
  \href {https://ui.adsabs.harvard.edu/abs/2016AJ....151...44D} {151, 44}

\bibitem[\protect\citeauthoryear{{Dekker}, {D'Odorico}, {Kaufer}, {Delabre}  \&
  {Kotzlowski}}{{Dekker} et~al.}{2000}]{Dekker:2000}
{Dekker} H.,  {D'Odorico} S.,  {Kaufer} A.,  {Delabre} B.,   {Kotzlowski} H.,
  2000, in {Iye} M.,  {Moorwood} A.~F.,  eds,  Society of Photo-Optical
  Instrumentation Engineers (SPIE) Conference Series Vol. 4008, Optical and IR
  Telescope Instrumentation and Detectors. pp 534--545,
  \mn@doi{10.1117/12.395512}

\bibitem[\protect\citeauthoryear{{Doughty}, {Hennawi}, {O{\~n}orbe}, {Davies}
  \& {Luki{\'c}}}{{Doughty} et~al.}{2023}]{2023arXiv231201480D}
{Doughty} C.~C.,  {Hennawi} J.~F.,  {O{\~n}orbe} J.,  {Davies} F.~B.,
  {Luki{\'c}} Z.,  2023, \mn@doi [arXiv e-prints] {10.48550/arXiv.2312.01480},
  \href {https://ui.adsabs.harvard.edu/abs/2023arXiv231201480D} {p.
  arXiv:2312.01480}

\bibitem[\protect\citeauthoryear{{Eisenstein}, {Seo}  \& {White}}{{Eisenstein}
  et~al.}{2007}]{Eisenstein2007}
{Eisenstein} D.~J.,  {Seo} H.-J.,   {White} M.,  2007, \mn@doi [\apj]
  {10.1086/518755}, \href
  {https://ui.adsabs.harvard.edu/abs/2007ApJ...664..660E} {664, 660}

\bibitem[\protect\citeauthoryear{{Farr} et~al.,}{{Farr} et~al.}{2020}]{Farr20}
{Farr} J.,  et~al., 2020, \mn@doi [\jcap] {10.1088/1475-7516/2020/03/068},
  \href {https://ui.adsabs.harvard.edu/abs/2020JCAP...03..068F} {2020, 068}

\bibitem[\protect\citeauthoryear{{Font-Ribera}, {McDonald}  \&
  {Slosar}}{{Font-Ribera} et~al.}{2018}]{Font-Ribera:2018}
{Font-Ribera} A.,  {McDonald} P.,   {Slosar} A.,  2018, \mn@doi [\jcap]
  {10.1088/1475-7516/2018/01/003}, \href
  {https://ui.adsabs.harvard.edu/abs/2018JCAP...01..003F} {2018, 003}

\bibitem[\protect\citeauthoryear{{Gaikwad}, {Srianand}, {Khaire}  \&
  {Choudhury}}{{Gaikwad} et~al.}{2019}]{Gaikwad:2019}
{Gaikwad} P.,  {Srianand} R.,  {Khaire} V.,   {Choudhury} T.~R.,  2019, \mn@doi
  [\mnras] {10.1093/mnras/stz2692}, \href
  {https://ui.adsabs.harvard.edu/abs/2019MNRAS.490.1588G} {490, 1588}

\bibitem[\protect\citeauthoryear{{Gaikwad}, {Srianand}, {Haehnelt}  \&
  {Choudhury}}{{Gaikwad} et~al.}{2021}]{Gaikwad:2021}
{Gaikwad} P.,  {Srianand} R.,  {Haehnelt} M.~G.,   {Choudhury} T.~R.,  2021,
  \mn@doi [\mnras] {10.1093/mnras/stab2017}, \href
  {https://ui.adsabs.harvard.edu/abs/2021MNRAS.506.4389G} {506, 4389}

\bibitem[\protect\citeauthoryear{{Garny}, {Konstandin}, {Sagunski}  \&
  {Viel}}{{Garny} et~al.}{2021}]{Garny:2021}
{Garny} M.,  {Konstandin} T.,  {Sagunski} L.,   {Viel} M.,  2021, \mn@doi
  [\jcap] {10.1088/1475-7516/2021/03/049}, \href
  {https://ui.adsabs.harvard.edu/abs/2021JCAP...03..049G} {2021, 049}

\bibitem[\protect\citeauthoryear{{Garzilli}, {Bolton}, {Kim}, {Leach}  \&
  {Viel}}{{Garzilli} et~al.}{2012}]{Garzilli:2012}
{Garzilli} A.,  {Bolton} J.~S.,  {Kim} T.~S.,  {Leach} S.,   {Viel} M.,  2012,
  \mn@doi [\mnras] {10.1111/j.1365-2966.2012.21223.x}, \href
  {https://ui.adsabs.harvard.edu/abs/2012MNRAS.424.1723G} {424, 1723}

\bibitem[\protect\citeauthoryear{{Garzilli}, {Magalich}, {Theuns}, {Frenk},
  {Weniger}, {Ruchayskiy}  \& {Boyarsky}}{{Garzilli}
  et~al.}{2019}]{Garzilli:2019}
{Garzilli} A.,  {Magalich} A.,  {Theuns} T.,  {Frenk} C.~S.,  {Weniger} C.,
  {Ruchayskiy} O.,   {Boyarsky} A.,  2019, \mn@doi [\mnras]
  {10.1093/mnras/stz2188}, \href
  {https://ui.adsabs.harvard.edu/abs/2019MNRAS.489.3456G} {489, 3456}

\bibitem[\protect\citeauthoryear{{Givans} \& {Hirata}}{{Givans} \&
  {Hirata}}{2020}]{Givans:2020}
{Givans} J.~J.,  {Hirata} C.~M.,  2020, \mn@doi [\prd]
  {10.1103/PhysRevD.102.023515}, \href
  {https://ui.adsabs.harvard.edu/abs/2020PhRvD.102b3515G} {102, 023515}

\bibitem[\protect\citeauthoryear{{Givans} et~al.,}{{Givans}
  et~al.}{2022}]{Givans:2022}
{Givans} J.~J.,  et~al., 2022, \mn@doi [\jcap] {10.1088/1475-7516/2022/09/070},
  \href {https://ui.adsabs.harvard.edu/abs/2022JCAP...09..070G} {2022, 070}

\bibitem[\protect\citeauthoryear{{Goldstein}, {Hill}, {Ir{\v{s}}i{\v{c}}}  \&
  {Sherwin}}{{Goldstein} et~al.}{2023}]{2023PhRvL.131t1001G}
{Goldstein} S.,  {Hill} J.~C.,  {Ir{\v{s}}i{\v{c}}} V.,   {Sherwin} B.~D.,
  2023, \mn@doi [\prl] {10.1103/PhysRevLett.131.201001}, \href
  {https://ui.adsabs.harvard.edu/abs/2023PhRvL.131t1001G} {131, 201001}

\bibitem[\protect\citeauthoryear{{Gordon} et~al.,}{{Gordon}
  et~al.}{2023a}]{Gordon:2023}
{Gordon} C.,  et~al., 2023a, \mn@doi [arXiv e-prints]
  {10.48550/arXiv.2308.10950}, \href
  {https://ui.adsabs.harvard.edu/abs/2023arXiv230810950G} {p. arXiv:2308.10950}

\bibitem[\protect\citeauthoryear{{Gordon} et~al.,}{{Gordon}
  et~al.}{2023b}]{2023JCAP...11..045G}
{Gordon} C.,  et~al., 2023b, \mn@doi [\jcap] {10.1088/1475-7516/2023/11/045},
  \href {https://ui.adsabs.harvard.edu/abs/2023JCAP...11..045G} {2023, 045}

\bibitem[\protect\citeauthoryear{{Greene}, {Bezanson}, {Ouchi}, {Silverman}  \&
  {the PFS Galaxy Evolution Working Group}}{{Greene} et~al.}{2022}]{2022PFSGE}
{Greene} J.,  {Bezanson} R.,  {Ouchi} M.,  {Silverman} J.,   {the PFS Galaxy
  Evolution Working Group} 2022, \mn@doi [arXiv e-prints]
  {10.48550/arXiv.2206.14908}, \href
  {https://ui.adsabs.harvard.edu/abs/2022arXiv220614908G} {p. arXiv:2206.14908}

\bibitem[\protect\citeauthoryear{{Herrera-Alcantar} et~al.,}{{Herrera-Alcantar}
  et~al.}{2023}]{2024arXiv240100303H}
{Herrera-Alcantar} H.~K.,  et~al., 2023, \mn@doi [arXiv e-prints]
  {10.48550/arXiv.2401.00303}, \href
  {https://ui.adsabs.harvard.edu/abs/2024arXiv240100303H} {p. arXiv:2401.00303}

\bibitem[\protect\citeauthoryear{{Horowitz}, {de Belsunce}  \&
  {Lukic}}{{Horowitz} et~al.}{2024}]{Horowitz:2024}
{Horowitz} B.,  {de Belsunce} R.,   {Lukic} Z.,  2024, \mn@doi [arXiv e-prints]
  {10.48550/arXiv.2403.17294}, \href
  {https://ui.adsabs.harvard.edu/abs/2024arXiv240317294H} {p. arXiv:2403.17294}

\bibitem[\protect\citeauthoryear{{Hui}, {Stebbins}  \& {Burles}}{{Hui}
  et~al.}{1999}]{Hui:1999}
{Hui} L.,  {Stebbins} A.,   {Burles} S.,  1999, \mn@doi [\apjl]
  {10.1086/311826}, \href
  {https://ui.adsabs.harvard.edu/abs/1999ApJ...511L...5H} {511, L5}

\bibitem[\protect\citeauthoryear{{Ir{\v s}i{\v c}} et~al.,}{{Ir{\v s}i{\v c}}
  et~al.}{2017}]{Irsic17}
{Ir{\v s}i{\v c}} V.,  et~al., 2017, preprint, \href
  {http://adsabs.harvard.edu/abs/2017arXiv170201764I} {} (\mn@eprint {arXiv}
  {1702.01764})

\bibitem[\protect\citeauthoryear{{Ir{\v{s}}i{\v{c}}} \&
  {McQuinn}}{{Ir{\v{s}}i{\v{c}}} \& {McQuinn}}{2018}]{Irsic:2018}
{Ir{\v{s}}i{\v{c}}} V.,  {McQuinn} M.,  2018, \mn@doi [\jcap]
  {10.1088/1475-7516/2018/04/026}, \href
  {https://ui.adsabs.harvard.edu/abs/2018JCAP...04..026I} {2018, 026}

\bibitem[\protect\citeauthoryear{{Ir{\v{s}}i{\v{c}}}, {Xiao}  \&
  {McQuinn}}{{Ir{\v{s}}i{\v{c}}} et~al.}{2020}]{Irsic:2020}
{Ir{\v{s}}i{\v{c}}} V.,  {Xiao} H.,   {McQuinn} M.,  2020, \mn@doi [\prd]
  {10.1103/PhysRevD.101.123518}, \href
  {https://ui.adsabs.harvard.edu/abs/2020PhRvD.101l3518I} {101, 123518}

\bibitem[\protect\citeauthoryear{{Ir{\v{s}}i{\v{c}}}
  et~al.,}{{Ir{\v{s}}i{\v{c}}} et~al.}{2023}]{Irsic:2023}
{Ir{\v{s}}i{\v{c}}} V.,  et~al., 2023, \mn@doi [arXiv e-prints]
  {10.48550/arXiv.2309.04533}, \href
  {https://ui.adsabs.harvard.edu/abs/2023arXiv230904533I} {p. arXiv:2309.04533}

\bibitem[\protect\citeauthoryear{{Ivanov}}{{Ivanov}}{2023}]{Ivanov:2023}
{Ivanov} M.~M.,  2023, \mn@doi [arXiv e-prints] {10.48550/arXiv.2309.10133},
  \href {https://ui.adsabs.harvard.edu/abs/2023arXiv230910133I} {p.
  arXiv:2309.10133}

\bibitem[\protect\citeauthoryear{{Kaiser}}{{Kaiser}}{1987}]{Kaiser1987}
{Kaiser} N.,  1987, \mn@doi [\mnras] {10.1093/mnras/227.1.1}, \href
  {https://ui.adsabs.harvard.edu/abs/1987MNRAS.227....1K} {227, 1}

\bibitem[\protect\citeauthoryear{{Kara{\c{c}}ayl{\i}}
  et~al.,}{{Kara{\c{c}}ayl{\i}} et~al.}{2024}]{2024MNRAS.tmp..176K}
{Kara{\c{c}}ayl{\i}} N.~G.,  et~al., 2024, \mn@doi [\mnras]
  {10.1093/mnras/stae171}, \href
  {https://ui.adsabs.harvard.edu/abs/2024MNRAS.tmp..176K} {}

\bibitem[\protect\citeauthoryear{Karaçaylı, Font-Ribera  \&
  Padmanabhan}{Karaçaylı et~al.}{2020}]{NaimQMLE2020}
Karaçaylı N.~G.,  Font-Ribera A.,   Padmanabhan N.,  2020, \mn@doi [Monthly
  Notices of the Royal Astronomical Society] {10.1093/mnras/staa2331}, 497,
  4742

\bibitem[\protect\citeauthoryear{Karim, Armengaud, Mention, Chabanier, Ravoux
  \& Lukić}{Karim et~al.}{2023}]{karim2023measurement}
Karim M. L.~A.,  Armengaud E.,  Mention G.,  Chabanier S.,  Ravoux C.,   Lukić
  Z.,  2023, Measurement of the small-scale 3D Lyman-$\alpha$ forest power
  spectrum (\mn@eprint {arXiv} {2310.09116})

\bibitem[\protect\citeauthoryear{Kirkby et~al.,}{Kirkby
  et~al.}{2021}]{david_kirkby_2021_4566008}
Kirkby D.,  et~al., 2021, desihub/specsim: August 2020 Release,
  \mn@doi{10.5281/zenodo.4566008}, \url
  {https://doi.org/10.5281/zenodo.4566008}

\bibitem[\protect\citeauthoryear{{Kobayashi}, {Murgia}, {De Simone},
  {Ir{\v{s}}i{\v{c}}}  \& {Viel}}{{Kobayashi} et~al.}{2017}]{Kobayashi:2017}
{Kobayashi} T.,  {Murgia} R.,  {De Simone} A.,  {Ir{\v{s}}i{\v{c}}} V.,
  {Viel} M.,  2017, \mn@doi [\prd] {10.1103/PhysRevD.96.123514}, \href
  {https://ui.adsabs.harvard.edu/abs/2017PhRvD..96l3514K} {96, 123514}

\bibitem[\protect\citeauthoryear{Lyke et~al.}{Lyke et~al.}{2020}]{eBOSS_DR16Q}
Lyke B.~W.,  et~al., 2020, \mn@doi [Astrophys. J. Suppl.]
  {10.3847/1538-4365/aba623}, 250, 8

\bibitem[\protect\citeauthoryear{McDonald}{McDonald}{2003}]{McDonald:2001}
McDonald P.,  2003, \mn@doi [Astrophys. J.] {10.1086/345945}, 585, 34

\bibitem[\protect\citeauthoryear{{McDonald} \& {Eisenstein}}{{McDonald} \&
  {Eisenstein}}{2007}]{McDonald:2007}
{McDonald} P.,  {Eisenstein} D.~J.,  2007, \mn@doi [\prd]
  {10.1103/PhysRevD.76.063009}, \href
  {https://ui.adsabs.harvard.edu/abs/2007PhRvD..76f3009M} {76, 063009}

\bibitem[\protect\citeauthoryear{{McDonald} et~al.,}{{McDonald}
  et~al.}{2006}]{McDonald06}
{McDonald} P.,  et~al., 2006, \mn@doi [\apjs] {10.1086/444361}, \href
  {https://ui.adsabs.harvard.edu/abs/2006ApJS..163...80M} {163, 80}

\bibitem[\protect\citeauthoryear{{McQuinn}}{{McQuinn}}{2016}]{McQuinn:2016}
{McQuinn} M.,  2016, \mn@doi [\araa] {10.1146/annurev-astro-082214-122355},
  \href {https://ui.adsabs.harvard.edu/abs/2016ARA&A..54..313M} {54, 313}

\bibitem[\protect\citeauthoryear{{Meiksin}}{{Meiksin}}{2009}]{Meiksin:2009}
{Meiksin} A.~A.,  2009, \mn@doi [Reviews of Modern Physics]
  {10.1103/RevModPhys.81.1405}, \href
  {https://ui.adsabs.harvard.edu/abs/2009RvMP...81.1405M} {81, 1405}

\bibitem[\protect\citeauthoryear{{Murgia}, {Ir{\v{s}}i{\v{c}}}  \&
  {Viel}}{{Murgia} et~al.}{2018}]{Murgia:2018}
{Murgia} R.,  {Ir{\v{s}}i{\v{c}}} V.,   {Viel} M.,  2018, \mn@doi [\prd]
  {10.1103/PhysRevD.98.083540}, \href
  {https://ui.adsabs.harvard.edu/abs/2018PhRvD..98h3540M} {98, 083540}

\bibitem[\protect\citeauthoryear{{Murgia}, {Scelfo}, {Viel}  \&
  {Raccanelli}}{{Murgia} et~al.}{2019}]{Murgia:2019}
{Murgia} R.,  {Scelfo} G.,  {Viel} M.,   {Raccanelli} A.,  2019, \mn@doi [\prl]
  {10.1103/PhysRevLett.123.071102}, \href
  {https://ui.adsabs.harvard.edu/abs/2019PhRvL.123g1102M} {123, 071102}

\bibitem[\protect\citeauthoryear{{Murphy}, {Kacprzak}, {Savorgnan}  \&
  {Carswell}}{{Murphy} et~al.}{2019}]{Murphy:2019}
{Murphy} M.~T.,  {Kacprzak} G.~G.,  {Savorgnan} G. A.~D.,   {Carswell} R.~F.,
  2019, \mn@doi [\mnras] {10.1093/mnras/sty2834}, \href
  {https://ui.adsabs.harvard.edu/abs/2019MNRAS.482.3458M} {482, 3458}

\bibitem[\protect\citeauthoryear{{Myers} et~al.,}{{Myers}
  et~al.}{2015}]{2015ApJS..221...27M}
{Myers} A.~D.,  et~al., 2015, \mn@doi [\apjs] {10.1088/0067-0049/221/2/27},
  \href {https://ui.adsabs.harvard.edu/abs/2015ApJS..221...27M} {221, 27}

\bibitem[\protect\citeauthoryear{{Noterdaeme} et~al.,}{{Noterdaeme}
  et~al.}{2012}]{2012A&A...547L...1N}
{Noterdaeme} P.,  et~al., 2012, \mn@doi [\aap] {10.1051/0004-6361/201220259},
  \href {https://ui.adsabs.harvard.edu/abs/2012A&A...547L...1N} {547, L1}

\bibitem[\protect\citeauthoryear{{O'Meara}, {Lehner}, {Howk}  \&
  {Prochaska}}{{O'Meara} et~al.}{2021}]{OMeara:2021}
{O'Meara} J.~M.,  {Lehner} N.,  {Howk} J.~C.,   {Prochaska} J.~X.,  2021,
  \mn@doi [\aj] {10.3847/1538-3881/abcbf2}, \href
  {https://ui.adsabs.harvard.edu/abs/2021AJ....161...45O} {161, 45}

\bibitem[\protect\citeauthoryear{{Palanque-Delabrouille}
  et~al.,}{{Palanque-Delabrouille} et~al.}{2013}]{PYB13}
{Palanque-Delabrouille} N.,  et~al., 2013, \mn@doi [\aap]
  {10.1051/0004-6361/201322130}, \href
  {https://ui.adsabs.harvard.edu/abs/2013A&A...559A..85P} {559, A85}

\bibitem[\protect\citeauthoryear{{Palanque-Delabrouille}, {Y{\`e}che},
  {Sch{\"o}neberg}, {Lesgourgues}, {Walther}, {Chabanier}  \&
  {Armengaud}}{{Palanque-Delabrouille} et~al.}{2020}]{Palanque2020}
{Palanque-Delabrouille} N.,  {Y{\`e}che} C.,  {Sch{\"o}neberg} N.,
  {Lesgourgues} J.,  {Walther} M.,  {Chabanier} S.,   {Armengaud} E.,  2020,
  \mn@doi [\jcap] {10.1088/1475-7516/2020/04/038}, \href
  {https://ui.adsabs.harvard.edu/abs/2020JCAP...04..038P} {2020, 038}

\bibitem[\protect\citeauthoryear{Pedersen, Font-Ribera, Rogers, McDonald,
  Peiris, Pontzen  \& Slosar}{Pedersen et~al.}{2021}]{Pedersen:2020}
Pedersen C.,  Font-Ribera A.,  Rogers K.~K.,  McDonald P.,  Peiris H.~V.,
  Pontzen A.,   Slosar A.,  2021, \mn@doi [JCAP]
  {10.1088/1475-7516/2021/05/033}, 05, 033

\bibitem[\protect\citeauthoryear{{Philcox}}{{Philcox}}{2021}]{Philcox:2021}
{Philcox} O. H.~E.,  2021, \mn@doi [\mnras] {10.1093/mnras/staa3882}, \href
  {https://ui.adsabs.harvard.edu/abs/2021MNRAS.501.4004P} {501, 4004}

\bibitem[\protect\citeauthoryear{{Philcox} \& {Eisenstein}}{{Philcox} \&
  {Eisenstein}}{2020}]{Philcox2020}
{Philcox} O. H.~E.,  {Eisenstein} D.~J.,  2020, \mn@doi [\mnras]
  {10.1093/mnras/stz3335}, \href
  {https://ui.adsabs.harvard.edu/abs/2020MNRAS.492.1214P} {492, 1214}

\bibitem[\protect\citeauthoryear{{Philcox} \& {Ivanov}}{{Philcox} \&
  {Ivanov}}{2022}]{2022PhRvD.105d3517P}
{Philcox} O. H.~E.,  {Ivanov} M.~M.,  2022, \mn@doi [\prd]
  {10.1103/PhysRevD.105.043517}, \href
  {https://ui.adsabs.harvard.edu/abs/2022PhRvD.105d3517P} {105, 043517}

\bibitem[\protect\citeauthoryear{{Pieri} et~al.,}{{Pieri}
  et~al.}{2016}]{2016sf2a.conf..259P}
{Pieri} M.~M.,  et~al., 2016, in {Reyl{\'e}} C.,  {Richard} J.,  {Cambr{\'e}sy}
  L.,  {Deleuil} M.,  {P{\'e}contal} E.,  {Tresse} L.,   {Vauglin} I.,  eds,
  SF2A-2016: Proceedings of the Annual meeting of the French Society of
  Astronomy and Astrophysics. pp 259--266 (\mn@eprint {arXiv} {1611.09388}),
  \mn@doi{10.48550/arXiv.1611.09388}

\bibitem[\protect\citeauthoryear{{Planck Collaboration}}{{Planck
  Collaboration}}{2020}]{Aghanim:2018eyx}
{Planck Collaboration} 2020, \mn@doi [\aap] {10.1051/0004-6361/201833910},
  \href {https://ui.adsabs.harvard.edu/abs/2020A&A...641A...6P} {641, A6}

\bibitem[\protect\citeauthoryear{{Planck Collaboration} et~al.,}{{Planck
  Collaboration} et~al.}{2016}]{2016A&A...594A..13P}
{Planck Collaboration} et~al., 2016, \mn@doi [\aap]
  {10.1051/0004-6361/201525830}, \href
  {https://ui.adsabs.harvard.edu/abs/2016A&A...594A..13P} {594, A13}

\bibitem[\protect\citeauthoryear{{Puchwein} et~al.,}{{Puchwein}
  et~al.}{2023}]{2023MNRAS.519.6162P}
{Puchwein} E.,  et~al., 2023, \mn@doi [\mnras] {10.1093/mnras/stac3761}, \href
  {https://ui.adsabs.harvard.edu/abs/2023MNRAS.519.6162P} {519, 6162}

\bibitem[\protect\citeauthoryear{{Ram{\'\i}rez-P{\'e}rez}, {Sanchez}, {Alonso}
  \& {Font-Ribera}}{{Ram{\'\i}rez-P{\'e}rez}
  et~al.}{2022}]{2022JCAP...05..002R}
{Ram{\'\i}rez-P{\'e}rez} C.,  {Sanchez} J.,  {Alonso} D.,   {Font-Ribera} A.,
  2022, \mn@doi [\jcap] {10.1088/1475-7516/2022/05/002}, \href
  {https://ui.adsabs.harvard.edu/abs/2022JCAP...05..002R} {2022, 002}

\bibitem[\protect\citeauthoryear{{Ravoux} et~al.,}{{Ravoux}
  et~al.}{2023}]{2023MNRAS.526.5118R}
{Ravoux} C.,  et~al., 2023, \mn@doi [\mnras] {10.1093/mnras/stad3008}, \href
  {https://ui.adsabs.harvard.edu/abs/2023MNRAS.526.5118R} {526, 5118}

\bibitem[\protect\citeauthoryear{{Rogers}, {Bird}, {Peiris}, {Pontzen},
  {Font-Ribera}  \& {Leistedt}}{{Rogers} et~al.}{2018}]{Rogers:2018}
{Rogers} K.~K.,  {Bird} S.,  {Peiris} H.~V.,  {Pontzen} A.,  {Font-Ribera} A.,
   {Leistedt} B.,  2018, \mn@doi [\mnras] {10.1093/mnras/sty603}, \href
  {https://ui.adsabs.harvard.edu/abs/2018MNRAS.476.3716R} {476, 3716}

\bibitem[\protect\citeauthoryear{{Rogers}, {Dvorkin}  \& {Peiris}}{{Rogers}
  et~al.}{2022}]{Rogers:2022}
{Rogers} K.~K.,  {Dvorkin} C.,   {Peiris} H.~V.,  2022, \mn@doi [\prl]
  {10.1103/PhysRevLett.128.171301}, \href
  {https://ui.adsabs.harvard.edu/abs/2022PhRvL.128q1301R} {128, 171301}

\bibitem[\protect\citeauthoryear{{Seljak}}{{Seljak}}{2012}]{Seljak:2012}
{Seljak} U.,  2012, \mn@doi [\jcap] {10.1088/1475-7516/2012/03/004}, \href
  {https://ui.adsabs.harvard.edu/abs/2012JCAP...03..004S} {2012, 004}

\bibitem[\protect\citeauthoryear{{Seljak} et~al.,}{{Seljak}
  et~al.}{2005}]{Seljak:2005}
{Seljak} U.,  et~al., 2005, \mn@doi [\prd] {10.1103/PhysRevD.71.103515}, \href
  {https://ui.adsabs.harvard.edu/abs/2005PhRvD..71j3515S} {71, 103515}

\bibitem[\protect\citeauthoryear{{Slosar} et~al.,}{{Slosar}
  et~al.}{2011}]{Slosar2011}
{Slosar} A.,  et~al., 2011, \mn@doi [\jcap] {10.1088/1475-7516/2011/09/001},
  \href {https://ui.adsabs.harvard.edu/abs/2011JCAP...09..001S} {2011, 001}

\bibitem[\protect\citeauthoryear{{Slosar} et~al.,}{{Slosar}
  et~al.}{2013}]{Slosar2013}
{Slosar} A.,  et~al., 2013, \mn@doi [\jcap] {10.1088/1475-7516/2013/04/026},
  \href {https://ui.adsabs.harvard.edu/abs/2013JCAP...04..026S} {2013, 026}

\bibitem[\protect\citeauthoryear{{Viel}, {Lesgourgues}, {Haehnelt}, {Matarrese}
   \& {Riotto}}{{Viel} et~al.}{2005}]{Viel:2005}
{Viel} M.,  {Lesgourgues} J.,  {Haehnelt} M.~G.,  {Matarrese} S.,   {Riotto}
  A.,  2005, \mn@doi [\prd] {10.1103/PhysRevD.71.063534}, \href
  {https://ui.adsabs.harvard.edu/abs/2005PhRvD..71f3534V} {71, 063534}

\bibitem[\protect\citeauthoryear{{Viel}, {Lesgourgues}, {Haehnelt}, {Matarrese}
   \& {Riotto}}{{Viel} et~al.}{2006}]{Viel:2006}
{Viel} M.,  {Lesgourgues} J.,  {Haehnelt} M.~G.,  {Matarrese} S.,   {Riotto}
  A.,  2006, \mn@doi [\prl] {10.1103/PhysRevLett.97.071301}, \href
  {https://ui.adsabs.harvard.edu/abs/2006PhRvL..97g1301V} {97, 071301}

\bibitem[\protect\citeauthoryear{{Viel}, {Haehnelt}  \& {Springel}}{{Viel}
  et~al.}{2010}]{Viel:2010}
{Viel} M.,  {Haehnelt} M.~G.,   {Springel} V.,  2010, \mn@doi [\jcap]
  {10.1088/1475-7516/2010/06/015}, \href
  {https://ui.adsabs.harvard.edu/abs/2010JCAP...06..015V} {2010, 015}

\bibitem[\protect\citeauthoryear{{Viel}, {Becker}, {Bolton}  \&
  {Haehnelt}}{{Viel} et~al.}{2013}]{Viel:2013}
{Viel} M.,  {Becker} G.~D.,  {Bolton} J.~S.,   {Haehnelt} M.~G.,  2013, \mn@doi
  [\prd] {10.1103/PhysRevD.88.043502}, \href
  {https://ui.adsabs.harvard.edu/abs/2013PhRvD..88d3502V} {88, 043502}

\bibitem[\protect\citeauthoryear{{Villasenor}, {Robertson}, {Madau}  \&
  {Schneider}}{{Villasenor} et~al.}{2022}]{Villasenor:2022}
{Villasenor} B.,  {Robertson} B.,  {Madau} P.,   {Schneider} E.,  2022, \mn@doi
  [\apj] {10.3847/1538-4357/ac704e}, \href
  {https://ui.adsabs.harvard.edu/abs/2022ApJ...933...59V} {933, 59}

\bibitem[\protect\citeauthoryear{{Villasenor}, {Robertson}, {Madau}  \&
  {Schneider}}{{Villasenor} et~al.}{2023}]{Villasenor:2023}
{Villasenor} B.,  {Robertson} B.,  {Madau} P.,   {Schneider} E.,  2023, \mn@doi
  [\prd] {10.1103/PhysRevD.108.023502}, \href
  {https://ui.adsabs.harvard.edu/abs/2023PhRvD.108b3502V} {108, 023502}

\bibitem[\protect\citeauthoryear{{Vogt} et~al.,}{{Vogt}
  et~al.}{1994}]{Vogt:1994}
{Vogt} S.~S.,  et~al., 1994, in {Crawford} D.~L.,  {Craine} E.~R.,  eds,
  Society of Photo-Optical Instrumentation Engineers (SPIE) Conference Series
  Vol. 2198, Instrumentation in Astronomy VIII. p.~362,
  \mn@doi{10.1117/12.176725}

\bibitem[\protect\citeauthoryear{{Walther}, {O{\~n}orbe}, {Hennawi}  \&
  {Luki{\'c}}}{{Walther} et~al.}{2019}]{Walther:2019}
{Walther} M.,  {O{\~n}orbe} J.,  {Hennawi} J.~F.,   {Luki{\'c}} Z.,  2019,
  \mn@doi [\apj] {10.3847/1538-4357/aafad1}, \href
  {https://ui.adsabs.harvard.edu/abs/2019ApJ...872...13W} {872, 13}

\bibitem[\protect\citeauthoryear{{Wilson}, {Ir{\v{s}}i{\v{c}}}  \&
  {McQuinn}}{{Wilson} et~al.}{2022}]{Wilson:2022}
{Wilson} B.,  {Ir{\v{s}}i{\v{c}}} V.,   {McQuinn} M.,  2022, \mn@doi [\mnras]
  {10.1093/mnras/stab3017}, \href
  {https://ui.adsabs.harvard.edu/abs/2022MNRAS.509.2423W} {509, 2423}

\bibitem[\protect\citeauthoryear{{York} et~al.,}{{York}
  et~al.}{2000}]{York2000}
{York} D.~G.,  et~al., 2000, \mn@doi [\aj] {10.1086/301513}, \href
  {http://adsabs.harvard.edu/abs/2000AJ....120.1579Y} {120, 1579}

\bibitem[\protect\citeauthoryear{{Zaldarriaga}}{{Zaldarriaga}}{2002}]{Zaldarriaga:2002}
{Zaldarriaga} M.,  2002, \mn@doi [\apj] {10.1086/324212}, \href
  {https://ui.adsabs.harvard.edu/abs/2002ApJ...564..153Z} {564, 153}

\bibitem[\protect\citeauthoryear{{de~Belsunce et al.}}{{de~Belsunce et
  al.}}{2024}]{Belsunce2024}
{de~Belsunce et al.} R.,  2024, in prep.

\bibitem[\protect\citeauthoryear{{de Jong} et~al.,}{{de Jong}
  et~al.}{2019}]{2019Msngr.175....3D}
{de Jong} R.~S.,  et~al., 2019, \mn@doi [The Messenger]
  {10.18727/0722-6691/5117}, \href
  {https://ui.adsabs.harvard.edu/abs/2019Msngr.175....3D} {175, 3}

\bibitem[\protect\citeauthoryear{{du Mas des Bourboux} et~al.,}{{du Mas des
  Bourboux} et~al.}{2020}]{dMdB:2020}
{du Mas des Bourboux} H.,  et~al., 2020, \mn@doi [\apj]
  {10.3847/1538-4357/abb085}, \href
  {https://ui.adsabs.harvard.edu/abs/2020ApJ...901..153D} {901, 153}

\makeatother
\end{thebibliography}

\appendix
\section{Survey Window Function for GRF} \label{sec:appendix_window_func}
\begin{figure}
    \centering
    \includegraphics[width=1\linewidth]{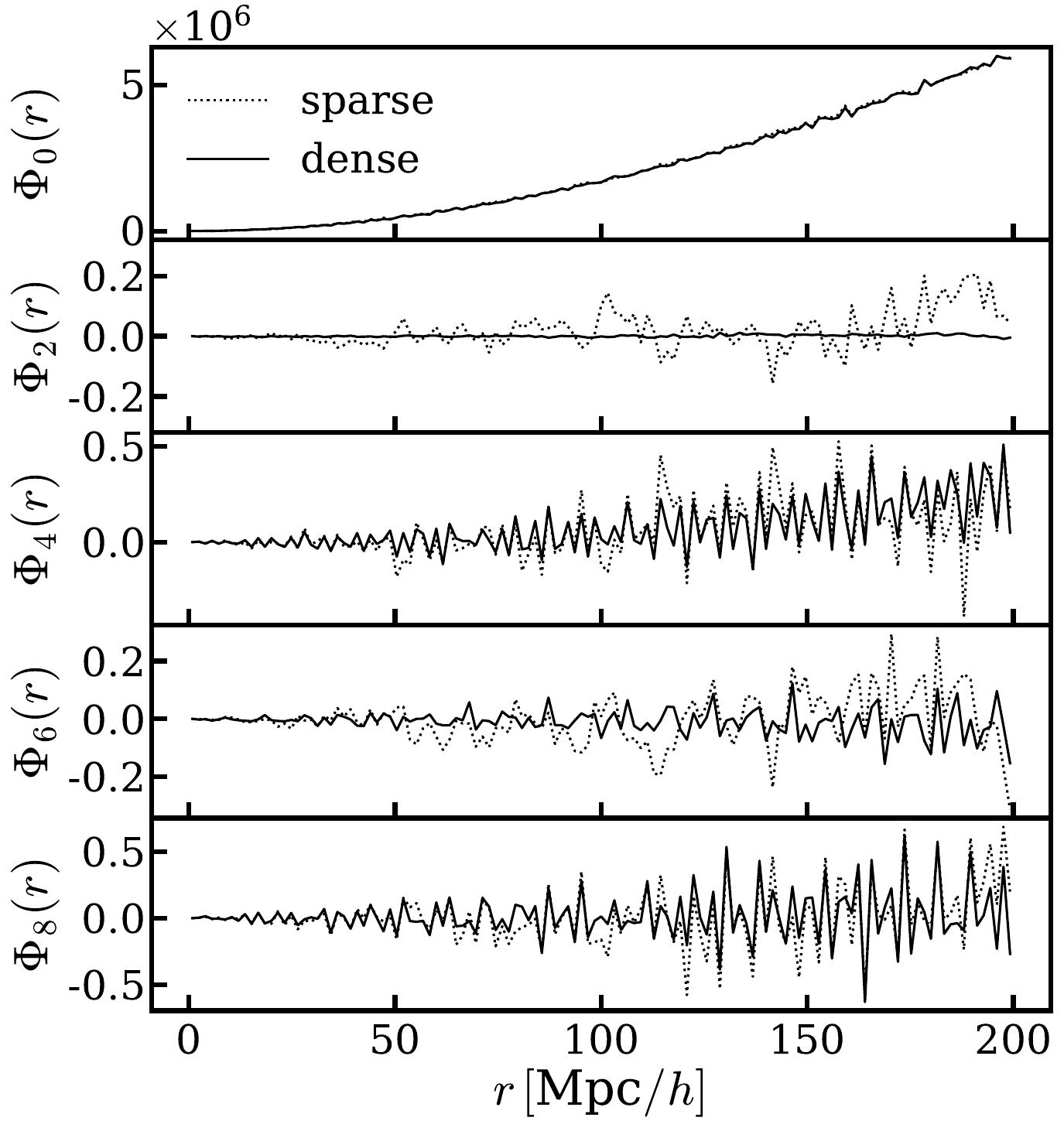}
    \caption{Window function of the data, $\Phi_{\ell}(r)$, for two different configurations discussed in Sec.~\ref{sec:GRF}: (i) sparse with $\sim 2\, \mathrm{qso}/\mathrm{deg}^2$; and (ii) dense with an eBOSS DR16-like sampling of $30\, \mathrm{qso}/\mathrm{deg}^2$. The dense window function has been normalized by its number density for ease of comparison.}
    \label{fig:phi-window}
\end{figure}

In Fig.~\ref{fig:phi-window} we show the multipoles of the window function, $\Phi_{\ell}(r)$, for a sparse and a dense survey configuration with $\sim 2-3$ and $30\, \mathrm{qso}/\mathrm{deg}^2$, respectively. The window function for the dense configuration is normalized by its number density to illustrate that reducing the number of skewers only increases the noise in each $r$ bin.

\begin{figure}
    \centering
    \includegraphics[width=1\linewidth]{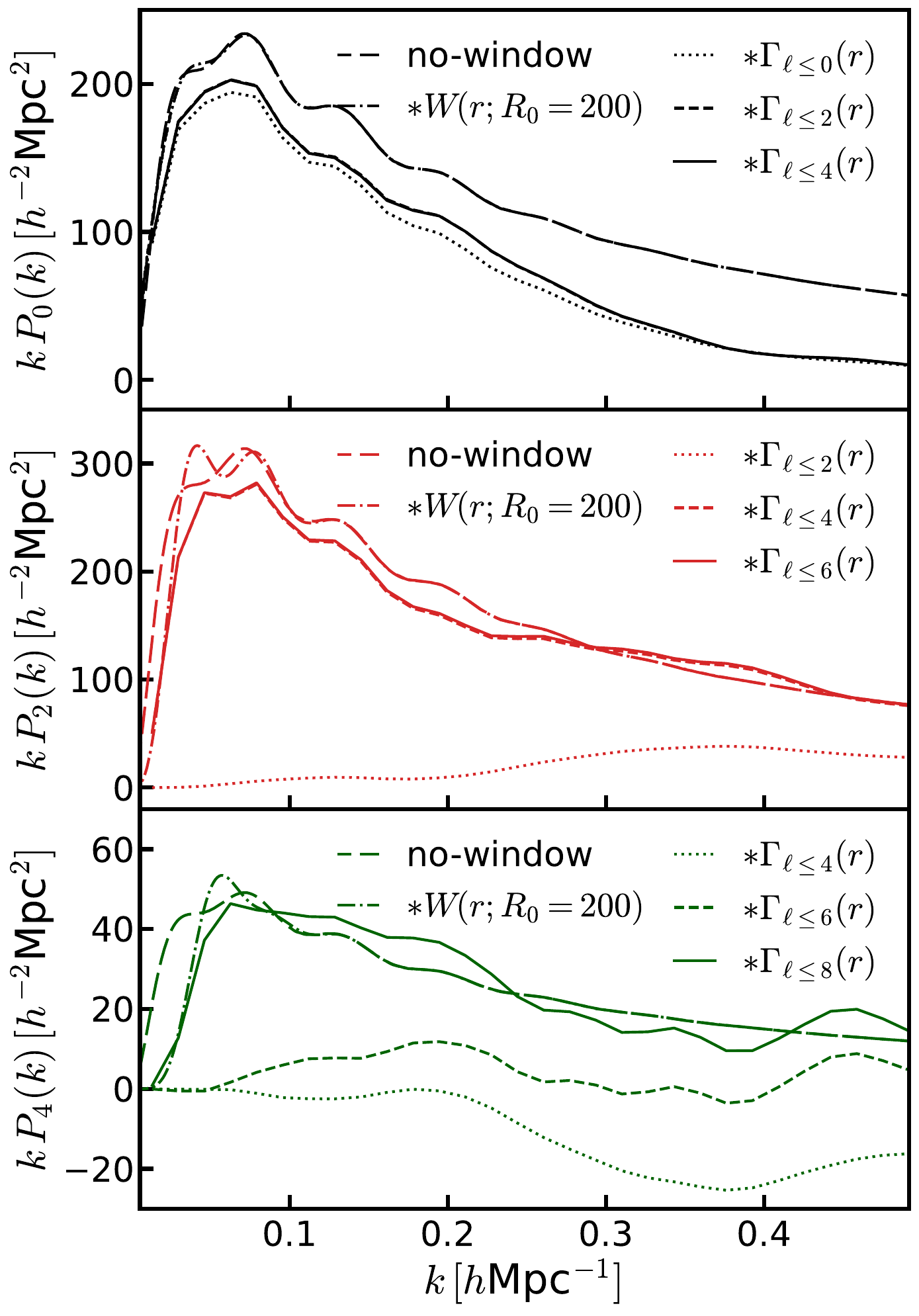}
    \vspace{-0.1in}
    \caption{Effect of the convolution of the theory power spectrum with the pair-count truncation window function $W(r;R_0)$ and the survey geometry $\Gamma_{\ell}^2 = \Phi_{\ell}(r,\mu)W(r;R_0)$ up to $\ell_{\rm max}$ evaluated at the same wavenumbers as for the data in Sec.~\ref{sec:p3d_eboss}. The power spectrum monopole contributions up to $\ell_{\rm max}=4$, quadrupole convolution up to $\ell_{\rm max}=6$ and hexadecapole convolution up to $\ell_{\rm max}=8$ are shown from top to bottom respectively.}
    \label{fig:pk_window}
    \vspace{-0.1in}
\end{figure}

In Fig.~\ref{fig:pk_window} we consider the impact of window function convolution on the linear theory power spectrum. We show five cases in the Figure: (i) The raw linear theory power spectrum; (ii) The convolution with the pair truncation window function with $R_0=200\hinvMpc$; (iii) The window-convolved spectra, defined in Eqs.~\eqref{eq:conv1}-\eqref{eq:conv3}, including only $\xi_0$ contributions; (iv) As (iii), but adding $\xi_2$ contributions; (v) As (iv) but adding $\xi_4$ contributions. Here, $\Gamma^2$ denotes the window matrix obtained from the survey geometry through pair-counting without the $\delta_F$ weights. It is interesting to note that the window pair separation function only affects very large scales below $k\simlt 0.08 \hMpcinv$; this is as expected since the $W(r;R_0)$ function given in Eq.~\eqref{eq:wofr} was chosen to reduce aliasing \citep{Philcox2020}. For the monopole spectra (top panel), we find large contributions from the monopole and quadrupole windows \textit{at all scales}, though negligible effects from the hexadecapole. In the middle panel we show the power spectrum quadrupole with contributions up to $\ell_{\rm max}=6$ where the hexadecapole contributions are strongest and higher order terms have vanishing power. In the bottom panel we show the theory convolved hexadecapole with window matrix contributions up to $\ell_{\rm max}=8$. Here it is important to note that all multipoles of the window matrix contribute visibly to the final window convolved power spectrum. This stems from the integrand of the Hankel transform from the window convolved correlation function to the power spectrum multipoles. Note that the integrands of the quadrupole and hexadecapole mix a wide range of scales; this is discussed in the context of the eBOSS DR16 data in Appendix \ref{sec:DM}.

\section{Forward modeling distortions from continuum fitting} \label{sec:DM}
\begin{figure}
    \centering
    \includegraphics[width=1\linewidth]{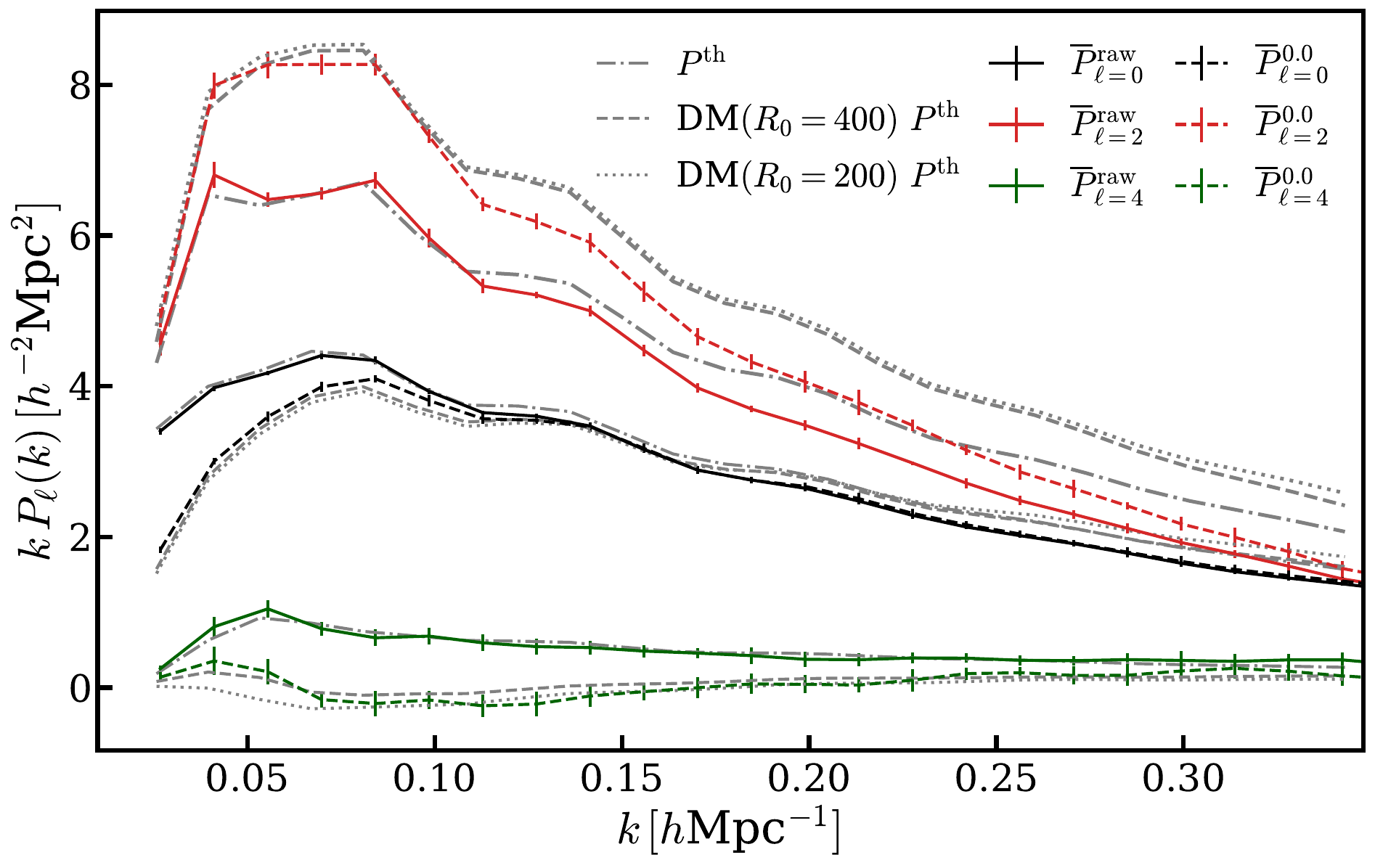}
    \vspace{-0.2in}
    \caption{Comparison of power spectra measured from `raw' (solid colored lines) and continuum-fitted `eboss-0.0' (dashed colored lines) \Lyacolore mocks shown for the monopole (black), quadrupole (red) and hexadecapole (green lines), respectively. The linear-theory Kaiser power spectrum, $P^{\rm th}$, shown as a dashed-dotted gray line fits well the large scales of the `raw' mocks. To illustrate the fidelity of the forward modeling of the distortion matrix, we show two variants of the analysis with distortion matrices computed for two different maximum separations: (i) $R_{\rm DM}=200 \hinvMpc$ shown as dotted line in gray; (ii) $R_{\rm DM}=400 \hinvMpc$ as dashed gray line.}
    \label{fig:DM_pk}
    \vspace{-0.1in}
\end{figure}

In Fig.~\ref{fig:DM_pk} we compare measured power spectra of `raw' (solid) and continuum-fitted `eboss-0.0' (dashed) mocks to a best-fit linear theory power spectrum. We apply the distortion matrix, denoted by DM, to the dashed-dotted linear power spectrum to test the range of validity of our distortion matrix treatment. The resulting distorted linear theory power is shown for two different maximum separations: (i) $R_{\rm DM}=200 \hinvMpc$ (dotted gray line) ; and (ii) $R_{\rm DM}=400 \hinvMpc$ (dashed). The continuum fitting approach introduced in Sec.~\ref{sec:cont_fit} suppresses modes along the line-of-sight, effectively mixing $\td$ fields resulting in distortions in the measured power spectra (or correlation functions). The resulting distortion matrix is computed from the `eboss-0.0' mocks using Eq.~\eqref{eq:DM}. 

When comparing the measured power spectra of the `raw' to the `eboss-0.0' mocks we find a suppression of the monopole on all scales which is most pronounced at $k\simlt 0.1 \hMpcinv$ by up to $\sim 20\%$. The quadrupole (red) is enhanced by $\sim 20\%$ on all scales. The hexadecapole (green) switches sign at the largest scales and is strongly damped. We compare the measured spectra to the linear-theory power spectra (gray lines) which are only valid on large scales (since the true power of these mocks is not known). Applying the DM to the multipoles, we find that the forward modeling of the distortion is accurate at the $\sim 5\%$ level. For the quadrupole and the hexadecapole we observe that the discrepancy between the distorted theory and the measurement for the `eboss-0.0' mocks is reduced at the largest scales by increasing the computed maximum separation of the distortion matrix from $200\hinvMpc$ to $400\hinvMpc$ in $4\hinvMpc$ bins in $\{\rpar, \rvperp\}$. We additionally tested that including higher order contributions of the theory correlation function, $\xi_{\ell}(r)$ up to $\ell_{\rm max}=6$, and including all the window matrix multipoles in Eqs.~\eqref{eq:conv1}-\eqref{eq:conv3}, did have a negligible effect on the theory prediction of the multipoles. This additional level of complexity can be removed from the analysis pipeline by using a continuum fitting method that does not introduce distortions, \eg, a PCA-based continuum fitting \citep{Belsunce2024} at the expense of obtaining noisier continuum estimates.

\begin{figure}
    \centering
    \includegraphics[width=1\linewidth]{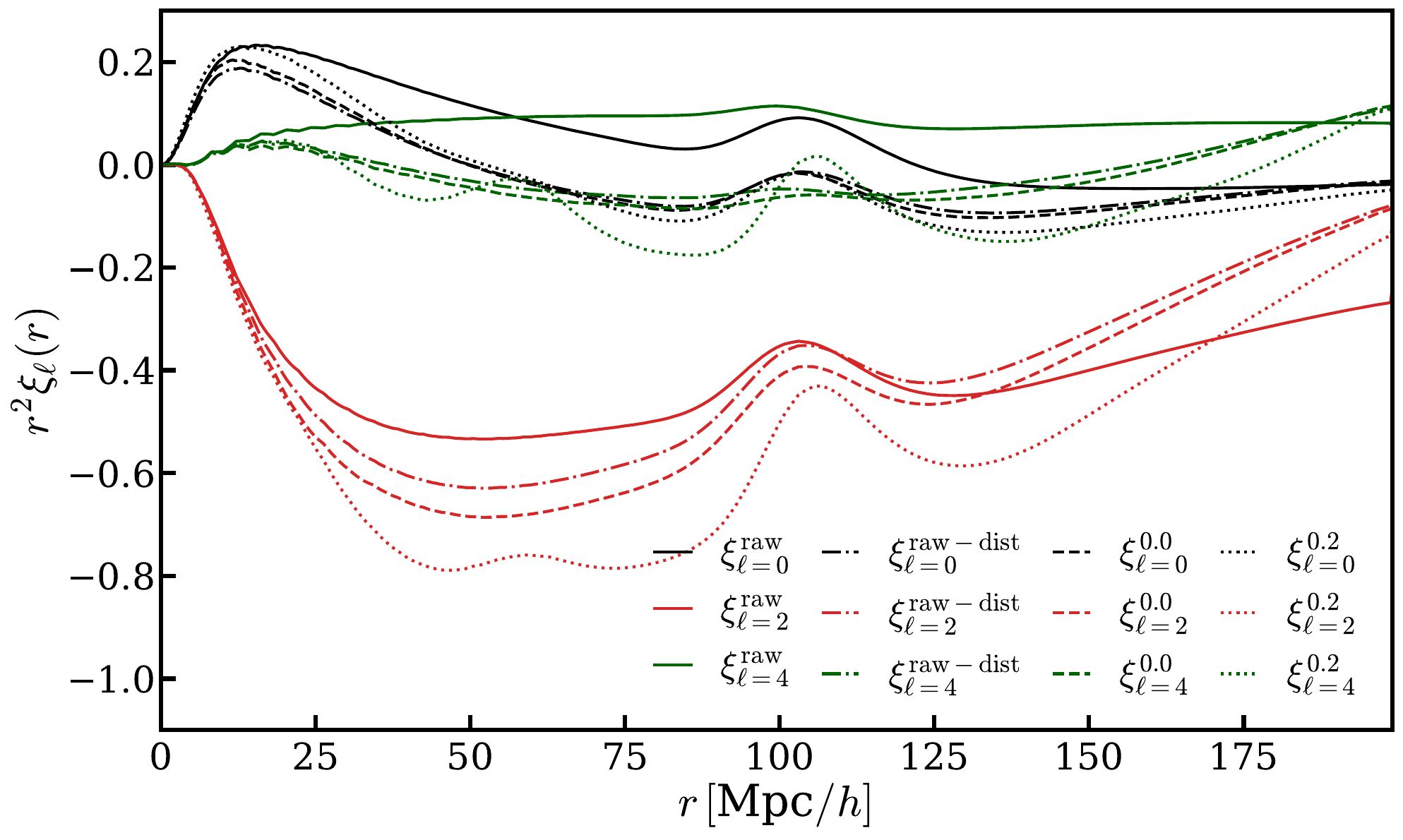}
    \vspace{-0.1in}
    \caption{Effect of the distortion matrix and contaminants on the theory correlation function multipoles: monopole (black), quadrupole (red) and hexadecapole (green lines), respectively. The solid lines are the best-fit correlation functions obtained from the mean correlation function measurements \citepalias{dMdB:2020} which agree well with the mean measured power spectrum for each mock configuration  in Fig.~\ref{fig:mocks_P3D}: solid lines correspond to the ones from `raw' mocks, dashed-dotted lines from `raw-dist' mocks including the mean subtraction displaying the effect of the distortion matrix, dashed lines from the `eboss-0.0' mocks which include the effect of noise and continuum fitting through the distortion matrix and the dotted lines are the correlation function multipoles of the `eboss-0.2' mocks which include contaminants and effects from continuum fitting.}
    \label{fig:DM_xi}
    \vspace{-0.1in}
\end{figure}

\begin{figure}
    \centering
    \includegraphics[width=1\linewidth]{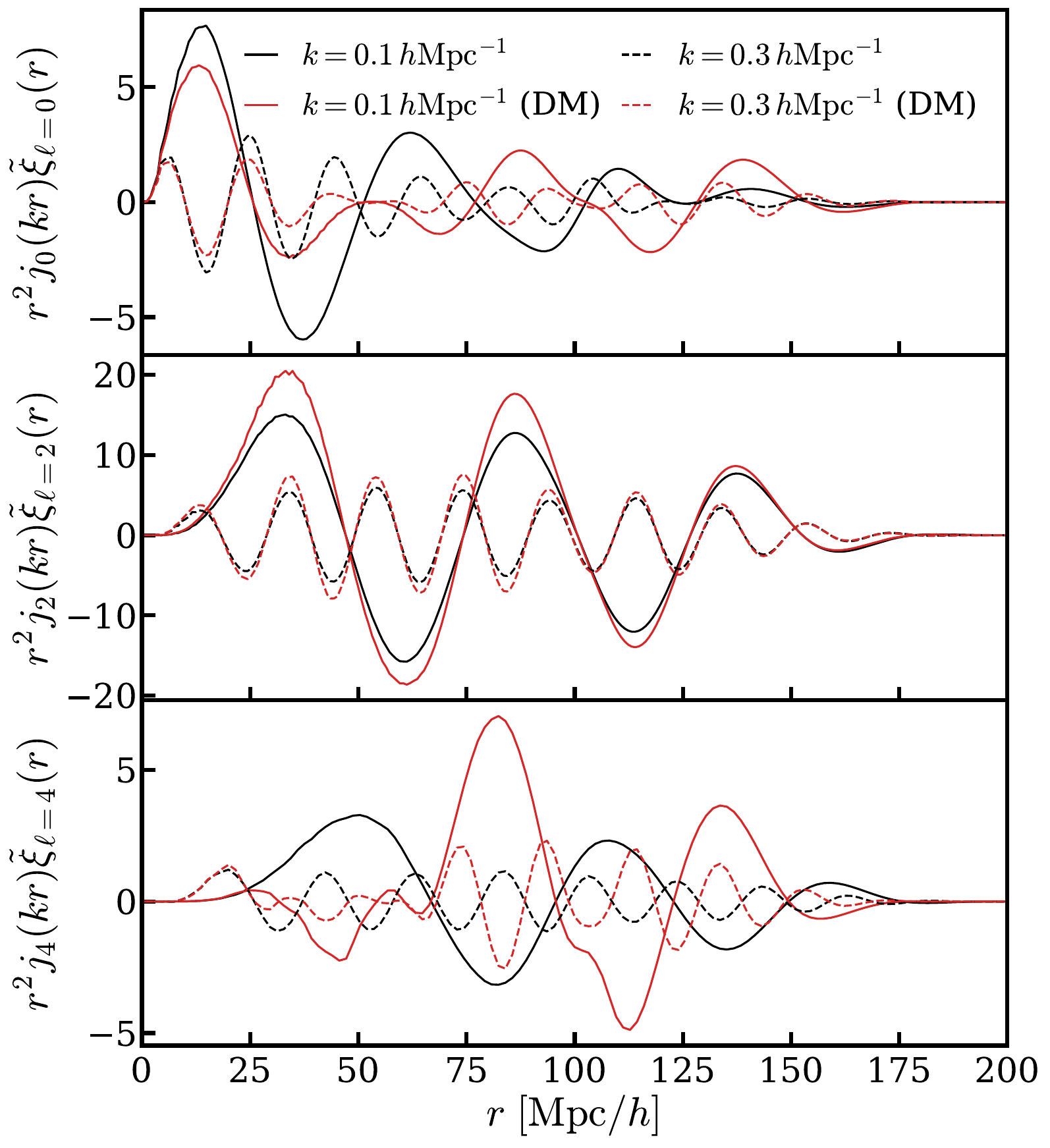}
    \vspace{-0.1in}
    \caption{Integrand of the Hankel transform given in Eq.~\eqref{eq:PWell} from the multipole of the correlation function to the corresponding multipole of the power spectrum (monopole top, quadrupole center, hexadecapole bottom). The pair truncation window $W(r;R_0=200\hinvMpc)$ is applied to the theory correlation functions. The solid (dashed) lines are for $k=0.1\hMpcinv$ ($k=0.3 \hMpcinv$); red (black) lines are with (with out) the distortion matrix.}
    \label{fig:xi2pk_Hankel}
    \vspace{-0.1in}
\end{figure}

For comparison we show the effect of the distortion matrix and contaminants on the multipoles of the theory correlation function in Fig.~\ref{fig:DM_xi} for the four \Lyacolore mock configurations. It is interesting to note that the application of the DM introduces a correlation between modes that are much further apart than a skewer length of a few hundred $\hinvMpc$'s, \ie in Fig.~\ref{fig:DM_pk} quadrupole modes at $k\sim 0.1 \hMpcinv$ and $k \sim 0.35 \hMpcinv$ are both affected by the DM. We conjecture that for a given mode $k$ the power spectrum quadrupole is sensitive to larger scales in the correlation function quadrupole than for the monopole. Therefore, we show in Fig.~\ref{fig:xi2pk_Hankel} the integrand of the Hankel transform of the window-convolved correlation function to the power spectrum multipoles for both wavenumbers $k\sim 0.1 \hMpcinv$ as a black solid and $k \sim 0.35 \hMpcinv$ as a dashed line, respectively. The application of the distortion matrix (black to red) mixes a wide range of scales for quadrupole and hexadecapole. 

\bsp    
\label{lastpage}
\end{document}